\documentclass[fleqn,usenatbib]{mnras}

\usepackage{newtxtext,newtxmath}

\usepackage[T1]{fontenc}
\usepackage{xspace}
\usepackage{comment}

\usepackage{tikz,xcolor,hyperref}
\definecolor{lime}{HTML}{A6CE39}
\DeclareRobustCommand{\orcidicon}{%
    \begin{tikzpicture}
    \draw[lime, fill=lime] (0,0) 
    circle [radius=0.16] 
    node[white] {{\fontfamily{qag}\selectfont \tiny ID}};
    \draw[white, fill=white] (-0.0625,0.095) 
    circle [radius=0.007];
    \end{tikzpicture}
    \hspace{-2mm}
}

\DeclareRobustCommand{\VAN}[3]{#2}
\let\VANthebibliography\thebibliography
\def\thebibliography{\DeclareRobustCommand{\VAN}[3]{##3}\VANthebibliography}

\DeclareRobustCommand{\DE}[3]{#2}
\let\DEthebibliography\thebibliography
\def\thebibliography{\DeclareRobustCommand{\DE}[3]{##3}\DEthebibliography}

\newcommand{\UC}{$\rm UNCOVER\_20466$\xspace}
\newcommand{\angstrom}{\textup{\AA}\xspace}

\newcommand{\lya}{$\rm Ly\alpha$\xspace}

\newcommand{\ciiiab}{$\rm CIII]\lambda\lambda 1907,1910$\xspace}

\newcommand{\mgiiab}{$\rm MgII\lambda\lambda 2796,2803$\xspace}
\newcommand{\oiia}{$\rm [OII]\lambda 3727$\xspace}
\newcommand{\oiib}{$\rm [OII]\lambda 3729$\xspace}
\newcommand{\oiiab}{$\rm [OII]\lambda\lambda 3727,3729$\xspace}

\newcommand{\neiiia}{$\rm [NeIII]\lambda3869$\xspace}
\newcommand{\neiiib}{$\rm [NeIII]\lambda3968$\xspace}
\newcommand{\neiiiab}{$\rm [NeIII]\lambda\lambda3869,3968$\xspace}
\newcommand{\oiiic}{$\rm [OIII]\lambda4363$\xspace}

\newcommand{\hg}{$\rm H\gamma$\xspace}
\newcommand{\hd}{$\rm H\delta$\xspace}
\newcommand{\hb}{$\rm H\beta$\xspace}
\newcommand{\oiiia}{[OIII]$\lambda4959$\xspace}
\newcommand{\oiiib}{[OIII]$\lambda5007$\xspace}
\newcommand{\oiiiab}{[OIII]$\lambda\lambda4959{,}5007$\xspace}

\newcommand{\niiab}{$\rm [NII]\lambda\lambda 6548,6584$\xspace}
\newcommand{\ha}{$\rm H\alpha$\xspace}

\newcommand{\siiab}{$\rm [SII]\lambda\lambda 6716,6731$\xspace}

\newcommand{\cii}{$\rm [CII]$\xspace}

\newcommand{\BH}{\rm BH}

\newcommand\textlcsc[1]{\textsc{\MakeLowercase{#1}}}

\newcommand{\pyneb}{\textlcsc{pyneb}\xspace}

\newcommand{\orcidGCJ}{\href{https://orcid.org/0000-0002-0267-9024}{\orcidicon}}
\newcommand{\orcidHU}{\href{https://orcid.org/0000-0003-4891-0794}{\orcidicon}}
\newcommand{\orcidRM}{\href{https://orcid.org/0000-0002-4985-3819}{\orcidicon}}
\newcommand{\orcidAJB}{\href{https://orcid.org/0000-0002-8651-9879}{\orcidicon}}
\newcommand{\orcidSCh}{\href{https://orcid.org/0000-0003-3458-2275}{\orcidicon}}
\newcommand{\orcidGC}{\href{https://orcid.org/0000-0002-5281-1417}{\orcidicon}}
\newcommand{\orcidFDE}{\href{https://orcid.org/0000-0003-2388-8172}{\orcidicon}}
\newcommand{\orcidKI}{\href{https://orcid.org/0000-0001-9840-4959}{\orcidicon}}
\newcommand{\orcidIJ}{\href{https://orcid.org/0009-0003-7423-8660}{\orcidicon}}
\newcommand{\orcidPGPG}{\href{https://orcid.org/0000-0003-4528-5639}{\orcidicon}}
\newcommand{\orcidMP}{\href{https://orcid.org/0000-0002-0362-5941}{\orcidicon}}
\newcommand{\orcidJS}{\href{https://orcid.org/0000-0001-6010-6809}{\orcidicon}}
\newcommand{\orcidST}{\href{https://orcid.org/0000-0002-8224-4505}{\orcidicon}}
\newcommand{\orcidXJ}{\href{https://orcid.org/0000-0002-1660-9502}{\orcidicon}}
\newcommand{\orcidSA}{\href{https://orcid.org/0000-0001-7997-1640}{\orcidicon}}
\newcommand{\orcidYI}{\href{https://orcid.org/0000-0002-5768-8235}{\orcidicon}}
\newcommand{\orcidGM}{\href{https://orcid.org/0009-0005-7383-6655}{\orcidicon}}
\newcommand{\orcidAM}{\href{https://orcid.org/0000-0002-9889-4238}{\orcidicon}}
\newcommand{\orcidRS}{\href{https://orcid.org/0000-0001-9317-2888}{\orcidicon}}
\newcommand{\orcidSCa}{\href{https://orcid.org/0000-0002-6719-380X}{\orcidicon}}

\usepackage{graphicx}	
\usepackage{amsmath}	

\title[BlackTHUNDER: UC20466]{BlackTHUNDER: Shedding light on a dormant and extreme little red dot at $\mathbf{z=8.50}$}

\author[G. C. Jones, et al.]{
\hspace{-1mm}Gareth C. Jones$^{1,2}$\orcidGCJ\thanks{E-mail: gj283@cam.ac.uk},
Hannah \"{U}bler$^{3}$\orcidHU,
Roberto Maiolino$^{1,2,4}$\orcidRM,
Xihan Ji$^{1,2}$\orcidXJ,
\newauthor
Alessandro Marconi $^{5,6}$\orcidAM,
Francesco D'Eugenio	$^{1,2}$\orcidFDE,
Santiago Arribas$^{7}$\orcidSA,
Andrew J. Bunker$^{8}$\orcidAJB,
\newauthor
Stefano Carniani$^{9}$\orcidSCa,
St\'ephane Charlot$^{10}$\orcidSCh,
Giovanni Cresci$^{11}$\orcidGC,
Kohei Inayoshi$^{12}$\orcidKI,
Yuki Isobe$^{1,2}$\orcidYI,
\newauthor
Ignas Juod{\v{z}}balis$^{1,2}$\orcidIJ,
Giovanni Mazzolari$^{3,13}$\orcidGM,
Pablo G. P\'erez-Gonz\'alez$^{7}$\orcidPGPG,
Michele Perna$^{7}$\orcidMP,
\newauthor
Raffaella Schneider$^{14}$\orcidRS,
Jan Scholtz$^{1,2}$\orcidJS,
Sandro Tacchella$^{1,2}$\orcidST
\\
$^{1}$Kavli Institute for Cosmology, University of Cambridge, Madingley Road, Cambridge CB3 0HA, UK\\
$^{2}$Cavendish Laboratory, University of Cambridge, 19 JJ Thomson Avenue, Cambridge CB3 0HE, UK\\
$^{3}$Max-Planck-Institut f\"ur extraterrestrische Physik, Gie{\ss}enbachstra{\ss}e 1, 85748 Garching, Germany\\
$^{4}$Department of Physics and Astronomy, University College London, Gower Street, London WC1E 6BT, UK\\
$^{5}$Dipartimento di Fisica e Astronomia, Università degli Studi di Firenze, Via G. Sansone 1,I-50019, Sesto Fiorentino, Firenze, Italy\\
$^{6}$INAF - Osservatorio Astrofisico di Arcetri, Largo E. Fermi 5, I-50125, Firenze, Italy\\
$^{7}$ Centro de Astrobiolog\'{i}a (CAB), CSIC-INTA, Ctra. de Ajalvir km 4, Torrej\'on de Ardoz, E-28850, Madrid, Spain\\
$^{8}$ Department of Physics, University of Oxford, Denys Wilkinson Building, Keble Road, Oxford OX1 3RH, UK\\
$^{9}$ Scuola Normale Superiore, Piazza dei Cavalieri 7, I-56126 Pisa, Italy\\
$^{10}$ Sorbonne Universit\'e, CNRS, UMR 7095, Institut d'Astrophysique de Paris, 98 bis bd Arago, 75014 Paris, France\\
$^{11}$ INAF - Osservatorio Astrofisco di Arcetri, largo E. Fermi 5, 50127 Firenze, Italy\\
$^{12}$ Kavli Institute for Astronomy and Astrophysics, Peking University, Beijing 100871, China\\
$^{13}$ INAF – Osservatorio di Astrofisica e Scienza dello Spazio di Bologna, Via Gobetti 93/3, I-40129 Bologna, Italy\\
$^{14}$ Dipartimento di Fisica, Sapienza Universit\'{a} di Roma, Piazzale Aldo Moro 5, 00185 Rome, Italy
}

\date{Accepted XXX. Received YYY; in original form ZZZ}

\pubyear{\the\year{}}

\begin{document}
\label{firstpage}
\pagerange{\pageref{firstpage}--\pageref{lastpage}}
\maketitle

\begin{abstract}
Recent photometric surveys with JWST have revealed a significant population of mysterious objects with red colours, compact morphologies, frequent signs of active galactic nucleus (AGN) activity, and negligible X-ray emission. 
These `Little Red Dots' (LRDs) have been explored through spectral and photometric studies, but their nature is still under debate. 
As part of the BlackTHUNDER survey, we have observed UNCOVER\_20466, one of the most distant LRDs known ($z=8.5$), with the JWST/NIRSpec IFU. 
Previous JWST/NIRCam and JWST/NIRSpec MSA observations of this source revealed its LRD nature, as well as the presence of an AGN. 
Using our NIRSpec IFU data, we confirm that UNCOVER\_20466 is an LRD (based on spectral slopes and compactness) that contains an overmassive black hole. 
However, our observed Balmer decrements do not suggest strong dust attenuation, resulting in a lower \hb-based bolometric luminosity and $\lambda_{\rm Edd}$ ($\sim10\%$) than previously found. 
This source lies on local relations between $M_{\rm BH}-\sigma_{*}$ and $M_{\rm BH}-M_{\rm dyn}$, suggesting that this could be a progenitor of the core of a lower-redshift galaxy.
We explore the possible evolution of this source, finding evidence for substantial black hole accretion in the past and a likely origin as a heavy seed at high redshift ($\sim10^3\,M_{\odot}$).
\lya emission is strongly detected, implying $f_{\rm esc}^{\rm Ly\alpha}\sim30\%$. 
The extremely high \oiiic/\hg ratio is indicative of not only AGN photoionization and heating, but also extremely high densities ($n_{\rm e}\sim 10^7\,\rm cm^{-3}$), suggesting that this black hole at such high redshift may be forming in an ultra-dense protogalaxy.
\end{abstract}




\section{Introduction}

The advent of JWST has revolutionised the field of high-redshift galaxy evolution, including the spectroscopic confirmation of galaxies at $z>10$ (within the first $\sim500$\,Myr of the Universe; e.g., \citealt{curt23,carn24,naid25b}), the construction of UV luminosity functions out to high redshift from photometric candidates (e.g., \citealt{donn24,robe24,whit25}), and new constraints on the progression and topology of ionisation of the Universe by galaxies in the epoch of reionisation (e.g., \citealt{ends24,simm24,whit25_lya}). One of the more surprising findings was the discovery of numerous compact, red objects in JWST/NIRCam data, which were given the name `little red dots' (LRDs; \citealt{matt24}). While a few red, compact sources were found to be local brown dwarfs \citep{lang23,hain24}, LRDs are primarily high-redshift galaxies (i.e., peaking in number between $4.5<z<8.0$; e.g., \citealt{koce25}). 

The red nature of these galaxies is usually caused by a distinctive `v'-shaped continuum from rest-optical to rest-UV, which is caused by red UV slopes ($\beta_{\rm UV}\sim-1.5$; compared to the bluer $\beta_{\rm UV}\sim-2.4$ found for star-forming galaxies; e.g., \citealt{saxe24_slope,topp24_slope,dott25}) and positive optical slopes ($\alpha_{\rm opt}>0$ where slopes are defined as $F_{\lambda}\propto\lambda^{\beta_{\rm UV}||\alpha_{\rm opt}}$; e.g, \citealt{gree24}). Their compactness is usually defined as a ratio of JWST/NIRCam photometric fluxes within different apertures (e.g., $C_{444W}\equiv f_{444}(0.4'')/f_{444}(0.2'')\lesssim1.7$; \citealt{koko24,akin25b,labb25}). In addition, many photometrically selected LRDs show broad hydrogen Balmer emission lines without similar broad emission in forbidden oxygen lines, which is robust evidence for an AGN nature (e.g., \citealt{furt23a,gree24,matt24,hvid25,koce25}). But unlike other AGN candidates, most LRDs are undetected in X-ray (e.g., \citealt{anan24, mazz24, maio25,  sacc25, yue24}) or submm/radio observations (e.g., \citealt{case25,mazz24_CEERS,oroz25,sett25,xiao25}, but cf. \citealt{fu25,golu25,rodr25}).

The nature of these sources is still under debate, as their properties are not trivially explained by common galactic models. This ambiguity is partially due to the fact that their natures can be described either by active galactic nucleus (AGN)-dominated galaxies or compact, dusty, star-forming galaxies (e.g., \citealt{bagg24}). Some investigations have looked into this by using different spectral energy distribution (SED) models: galaxy-only, AGN-only, and a combination of an AGN and host galaxy (hybrid; e.g., \citealt{koko23,trip24,leun25,rona25}). While these models can all fit the observed photometry of LRDs, it is difficult to robustly constrain the contribution of the AGN to the photometry of these sources. Therefore, the hybrid models return significantly lower stellar mass and dust attenuation ($A_{\rm V}$; e.g., \citealt{carr25}), but with high uncertainty on both physical quantities.


Previous works have suggested that LRDs represent black holes (either single or binary systems) embedded in massive gaseous envelopes (e.g., \citealt{inay25,kido25,lin25_env,ji25,naid25b,bege26}), possibly in the earliest phase of accretion (\citealt{inay25b,maio25_QSO1}), while others explore whether their properties could be explained by self-interacting dark matter halos (e.g., \citealt{robe25,feng25}), super-Eddington accretion (e.g., \citealt{liu25}), or population III supermassive stars (e.g., \citealt{nand25,jock26}).

Among the most extreme LRDs is the source \UC ($\alpha=3.640409$, $\delta=-30.386437$, $z=8.50$), which was observed with JWST/NIRCam as part the JWST Cycle 1 Treasury program UNCOVER (`Ultra-deep NIRCam and NIRSpec Observations Before the Epoch of Reionization'; PID 2561; PIs: I. Labbe, R. Bezanson; \citealt{beza24}). \citet{labb25} searched these data for objects with red colours (based on F115W to F444W colours) and compact morphology ($f_{\rm F444W}(r<0.2'')/f_{\rm F444W}(r<0.1'')<1.7$) that were well-detected in F444W ($\rm S/N>14$). This resulted in a sample of 26 reddened AGN candidates out of a parent sample of $5\times10^4$ sources. \UC was included as an AGN candidate\footnote{\UC was given the ID 13556 in \citet{labb25}.} with a very compact morphology ($r_{\rm e}=0.026''$) indistinguishable from the point spread function (PSF). Although \UC is gravitationally lensed by the Abell 2744 galaxy cluster, \citet{furt23b} find that \UC is only gravitationally magnified by a factor $\mu=1.33_{-0.02}^{+0.01}$. The best-fit photometric redshift was quite low ($z_{\rm phot}=5.17$), resulting in skewed best-fit parameters (e.g., $M_{\rm UV}$, $M_*$). The properties of \UC appeared to qualify it for the classification of an LRD, and \citet{lang23} confirmed that the LRD nature of \UC represents an extragalactic object rather than a brown dwarf.

\UC was then observed with the JWST/NIRSpec MSA, using the PRISM/CLEAR disperser/filter combination (spectral resolving power $R\sim100$) and an exposure time of 2.7\,hr. By analysing the resulting spectrum, \citet{koko23} found a higher spectroscopic redshift ($z_{\rm spec}=8.50$, resulting in $M_{\rm UV}=-18.11$; \citealt{fuji24}) and strong evidence for the presence of a broad-line AGN (see also analysis of \citealt{gree24}). A fit of the \hb line revealed a narrow and broad component, where $FWHM_{\rm broad}=3439\pm413$\,km\,s$^{-1}$. This broad component is not present in the well-detected \oiiiab lines, implying that it originates from a high-density broad line region (BLR) of an AGN (e.g., \citealt{bask05}). \UC is also classified as an AGN based on the narrow emission line ratio diagnostics proposed by \citet{mazz24}.

After correcting for lensing magnification, \citet{koko23} used the relations of \citet{gree05} to derive $\log_{10}(M_{\BH}/M_{\odot})=8.17\pm0.42$ (from broad \hb luminosity and width) and $\log_{10}(M_{\BH}/M_{\odot})=8.01\pm0.40$ (from a calibration based on $L_{5100}$). Assuming the conservative case of a stellar-only continuum (i.e., no AGN), \citet{koko23} determine $\log_{10}(M_{\rm *}/M_{\odot})<8.7$ (corrected for gravitational lensing). The bolometric luminosity is estimated as $L_{\rm bol}=(6.6\pm3.1)\times10^{45}$\,erg\,s$^{-1}$, using the \hb-based relation of \citet{ster12}. 

Together, this suggests that \UC features an extremely overmassive black hole, undergoing rapid accretion ($\lambda_{\rm Edd}\sim40\%$). This was further explored by \citet{zhanH25}, who used the empirical model \textlcsc{trinity} \citep{zhan23} to show that the black hole of \UC is overmassive for its redshift, similar to other high-redshift galaxies detected with JWST (e.g., UHZ1; \citealt{nata24}). While many other overmassive black holes have been discovered at $z>5$ (e.g., \citealt{goul23,uble23,furt24,maio24,maio25_QSO1,akin25,deug25,naid25,napo25,juod25,rina25}), \UC remains among the earliest ($z=8.5$, or $\sim600$\,Myr after the Big Bang) and most overmassive ($M_{\rm BH}/M_*\gtrsim0.06$) black holes in the observable Universe.

In this work, we present new JWST/NIRSpec IFU observations of \UC, taken as part of the BlackTHUNDER survey. In Section \ref{obs_sec}, we detail the JWST/NIRSpec IFU data used for this work, as well as archival JWST/NIRCam data. Section \ref{specsec} is an overview of our spectral extraction and fitting routine. We use the best-fit values to estimate several properties of \UC in Section \ref{analysis_sec}, and discuss the nature and evolution of this source in Section \ref{disc_sec}. We conclude in Section \ref{conc_sec}. 

We assume a concordance cosmology throughout, with H$_{\rm o}=70$\,km\,s$^{-1}$\,Mpc$^{-1}$, $\Omega_{\rm m}=0.3$, and $\Omega_{\rm \Lambda}=0.7$. At the redshift of \UC ($z=8.50$), $1''$ corresponds to $4.63$\, proper kpc (pkpc). Combining this with the gravitational magnification of \UC ($\mu=1.33$; \citealt{furt23b}), $1''$ in the image plane therefore corresponds to approximately 4.02\,pkpc in the source plane. Emission lines are named
based on their air wavelength, while we use their vacuum wavelengths for analysis (e.g. $\lambda_{\rm rest,[OIII]\lambda 5007}=5008.24\,\angstrom$).

\section{Observations and data reduction}\label{obs_sec}

\subsection{JWST/NIRSpec IFU}
The JWST/NIRSpec IFU data studied in this work originate from the BlackTHUNDER Large Programme (Black holes in THe early
Universe aNd their DensE surRoundings; PID 5015; PIs: H. Übler,
R. Maiolino). \UC was observed on 4-5 December, 2024 in PRISM/CLEAR (hereafter R100, $R\sim30-300$, $0.60{\,\rm \mu m}<\lambda_{\rm obs}<5.30{\,\rm \mu m}$) and G395H/F290LP (hereafter R2700, $R\sim2000-3700$, $2.87{\,\rm \mu m}<\lambda_{\rm obs}<5.14{\,\rm \mu m}$) using a 14-point medium cycling dither pattern (see Table \ref{obsdet} for more observation details).

\begin{table}
\centering
\begin{tabular}{c|ccc}
Disperser/Filter	&	Readout	&	Groups/Int	&	On-source	\\
	&	Pattern	&		&	Time [ks]	\\ \hline
G395H/F290LP	&	NRSIRS2	&	26	&	26.8	\\
PRISM/CLEAR	&	NRSIRS2RAPID	&	34	&	7.2	
\end{tabular}
\caption{BlackTHUNDER NIRSpec-IFU observation properties.}
\label{obsdet}
\end{table}

The raw data from these observations were downloaded from the the Barbara A. Mikulski Archive for Space Telescopes (MAST\footnote{\url{https://mast.stsci.edu/portal/Mashup/Clients/Mast/Portal.html}}). Calibration was performed using a customised version of the official pipeline (v1.15.0; \citealt{bush23}; see \citealt{pern23} for more details of customisation) with CRDS context 1293. These customisations account for 1/f noise correction, the subtraction of median values from count rate maps, manually masked cosmic rays and snowballs, open MSA slits, and rejection of artifacts identified by high flux-normalised derivatives along the dispersion direction of the count rate maps (\citealt{deug24}, using a rejection threshold of higher than $99.5\%$ of the resulting distribution for R100, and $98\%$ for R2700). By also utilising our multiple dithers and the drizzle algorithm (\citealt{fruc02}), we create two calibrated data cubes with spaxels of $0.05''$. We also removed pixels associated with artifacts by calculating the standard deviation of each spectral channel of the ERR extension (sigma\_clipped\_stats from astropy.stats).

Next, we performed background subtraction on each cube using the Background2D task of photutils \citep{brad21}. First, the spatial region of \UC was masked in each cube, in order to exclude the signal from the background estimation. We then use Background2D to divide the field of view into boxes of width $10$\,px ($0.5$\,arcsec), calculate the sigma-clipped noise level in each box ($3\sigma$), spatially smooth the resulting background cube using a square filter of $5$\,px ($0.25$\,arcsec), and apply a median filter of width 25 spectral pixels to find a final background cube. We inspected both background cubes to ensure that no narrow features were present, and used them to create background-subtracted cubes. 

\subsection{JWST/NIRCam}\label{NS_sec}
\UC was observed with NIRCam as part of two programs: UNCOVER and `Medium Bands, Mega Science' (PID 4111; PI: K. Suess; \citealt{sues24}). Combined, these surveys imaged \UC in 20 different filters (see Appendix \ref{filterapp} for details of each filter used in this work). We downloaded all available JWST/NIRCam images with stage 3 calibration from the MAST archive.

\section{Spectral analysis}\label{specsec}

Due to the very compact nature of \UC with respect to the JWST/NIRSpec IFU PSF, we do not spatially resolve the emission. Therefore, our primary analysis is focused on integrated spectra from the R100 and R2700 data cubes. While the field also contains a serendipitous source at lower redshift (`BlackBolt-1', see Appendix \ref{sere_sec}), we defer detailed analysis of this other object to a future work. 

\subsection{Spectral extraction}

The extraction aperture for \UC  should be carefully considered, as it must be small enough to maximise the S/N of each line, but large enough to contain all significant emission. Because \UC is very compact, it may be considered a point source for our observations, so the observed morphology will be dictated by the PSF. After a thorough exploration of the PSF of our data (see Appendix \ref{psfsec}), we proceed with an aperture that captures $60\%$ of the flux of the PSF at the observed wavelength of \oiiib, or a circular aperture of radius $0.125''$ centred on the observed emission.

\begin{figure}
    \centering
    \includegraphics[trim={0 2.5mm 0 0},clip,width=0.5\textwidth]{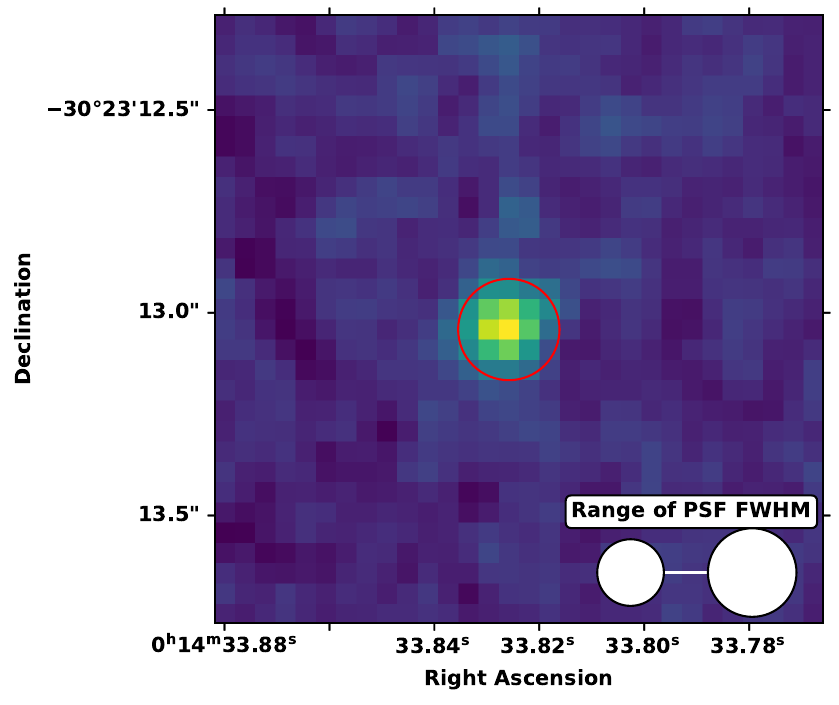}
    \caption{Flux map of \oiiib in \UC, compared to adopted aperture (red circle) and PSF FWHM range of our data (white circles).}
    \label{spat_fig}
\end{figure}

To visualise this aperture, we first create a flux map of \oiiib by summing the spectral channels of this line ($4.738<\lambda_{\rm obs}/\mu{\rm m}<4.793$) in our R100 data cube, creating a similar continuum map by summing adjacent, line-free spectral channels, and subtracting the two to create a line-only map (Figure \ref{spat_fig}; e.g., \citealt{scho24}). We compare the spatial distribution to both our aperture (red circle) and the range of PSF FWHM values in our data as found with \textlcsc{stpsf} (white circles, see Appendix \ref{psfsec}). Note that since the PSF at the wavelengths of other lines (e.g., \lya, \hb) are more compact, we will capture a higher percentage of these fluxes (see Appendix \ref{ALC_sec} for details of this aperture loss correction).

We extract spectra from both the R100 and R2700 data cubes using this aperture. Each spectrum is corrected for gravitational magnification (i.e., $\mu=1.33$, as assumed by \citealt{koko23}) and aperture loss.  As noted by other works (e.g., \citealt{uble23,uble24}), the ERR extension of each data cube contains the wavelength-dependent noise variation, but underestimates the true uncertainty. To correct for this, we follow the standard approach of determining the RMS scatter in a spectral range with no significant emission lines and scaling the error spectrum from ERR to feature the same scatter. 

\subsection{Spectral fitting}\label{sfsec}

Due to the availability of both R100 and R2700 data, we perform spectral fitting in two stages. First, we consider the data bluewards of the Balmer limit ($\lambda_{\rm rest}<0.3645\,\mu$m; Section \ref{bluefitsec}), or the location where many LRDs have been observed to feature a change in the spectral slope continuum, resulting in a `v'-shape (e.g., \citealt{hvid25,naid25,sett24}). This wavelength range is only covered by the R100 data. Next, we consider the data redwards of the Balmer limit, and fit the R100 and R2700 data simultaneously (Section \ref{redfitsec}).

\subsubsection{Blue fit}\label{bluefitsec}

First, we consider R100 data between $1050\,\angstrom\lesssim\lambda_{\rm rest}\lesssim3645\,\angstrom$ ($1.00\,\mu{\rm m}<\lambda_{\rm obs}<3.46\,\mu{\rm m}$). We note that while this data cube contains data at $\lambda_{\rm obs}<1.00\,\mu{\rm m}$, it is excluded due to high noise levels (as seen in the ERR extension of the data).

While the majority of the continuum emission may be described by a power law model with spectral slope $\beta_{\rm UV}$, the spectral slope here may deviate from the previously measured $\beta_{\rm UV}$ due to contributions from two-photon continuum or damped \lya clouds (e.g., \citealt{came24,tacc24}). In addition, this spectral region contains the \lya break which, when observed with the low spectral resolution of the PRISM disperser, may result in an artificial \lya damping wing (e.g., \citealt{hein24}). Thus, we follow the approach of other works (e.g., \citealt{jone25a}) to model the continuum here, as discussed below. 

An initial high-resolution model spectrum is created (channels of width $0.005\,\mu$m), and all spectral channels below the observed \lya wavelength ($\lambda_{\rm obs}<(1+z_{\rm Ly\alpha})\times 0.1216\,\mu$m) are set to a constant value\footnote{Ideally, this continuum value below the \lya break should be zero. However, we allow it to vary to account for calibration or imaging artifacts.}. The continuum between the \lya break and $\lambda_{\rm rest}=1500\angstrom$ is modelled as a power law ($F_{\lambda}\propto \lambda_{\rm obs}^{\beta_{\rm Ly\alpha}}$), and \lya flux is added to the first channel redwards of the \lya break. The continuum between $\lambda_{\rm rest}=1500\angstrom$ and the Balmer break is modelled as a separate power law ($F_{\lambda}\propto \lambda_{\rm obs}^{\beta_{\rm UV}}$), where we enforce continuity at $\lambda_{\rm rest}=1500\angstrom$.

The resulting continuum and \lya model is convolved with the wavelength-dependent line spread function (LSF\footnote{We determine the LSF using the fiducial resolving power curves recorded at \url{https://jwst-docs.stsci.edu/jwst-near-infrared-spectrograph/nirspec-instrumentation/nirspec-dispersers-and-filters}.}) and rebinned to match the wavelength bins of the observed data. We then add model Gaussians representing \ciiiab and \mgiiab, where the intrinsic widths and redshifts of each line are set to be equal. Due to the spectral resolution of the R100 data, we model each line pair as a single Gaussian. We account for the effect of the LSF by adding the intrinsic linewidth and LSF in quadrature. The free parameters of this approach are the continuum parameters (constant value bluewards of \lya break, value at $\lambda_{\rm rest}=1500\angstrom$, $\beta_{\rm Ly\alpha}$, $\beta_{\rm UV}$), line fluxes (\lya, \ciiiab and \mgiiab), and intrinsic linewidths ($FWHM_{\rm N}$). 

Previous studies of high-redshift galaxies with JWST/NIRSpec have detected strong emission from several rest-UV lines, including CIV$\lambda\lambda1548,1551$, NIV]$\lambda1486$, and NIII]$\lambda1747-1754$ (e.g., \citealt{bunk23,scho25_N,naid25b}). While we do not detect these lines in our data, it is possible that they are spread over multiple spectral channels by the wide R100 LSF. To avoid this effect biasing our measurement of the UV spectral slope, we exclude spectral channels within $10^4$\,km\,s$^{-1}$ (corresponding to approximately the FWHM of the LSF at these wavelengths) of each of these lines while fitting the data (see shaded regions in Figure \ref{gfit100}).

We use LMFIT \citep{newv14} in `least\_squares' mode to fit the extracted R100 spectrum, using the inverse variance as a weight. Due to the low spectral resolution of the R100 data, we fixed the redshift of each line and the \lya break to be $z=8.50$. Additionally, we fix the intrinsic linewidths to be $250$\,km\,s$^{-1}$. Our fits suggest that \mgiiab is undetected, so we set the flux of this line doublet to 0 in the model and use the error spectrum to derive an upper limit on its flux. The best-fit parameters are listed in Table \ref{linefluxtable}.

\subsubsection{Red fit}\label{redfitsec}

Next, we consider data between $3645\,\angstrom\lesssim\lambda_{\rm rest}\lesssim5475\,\angstrom$ ($3.46\,\mu{\rm m}<\lambda_{\rm obs}<5.2\,\mu{\rm m}$)\footnote{Similarly to above, we exclude $\lambda_{\rm obs}>5.20\,\mu{\rm m}$ data due to artifacts created during background subtraction.}. This range is covered by both the R100 and R2700 data, so we fit a model to both spectra simultaneously.

The continuum is modelled as a single power-law component ($F_{\lambda}\propto\lambda_{\rm obs}^{\alpha_{\rm opt}}$), and we include emission from 
\oiiab, \neiiiab, \hd, \hg, \oiiic, \hb, and \oiiiab. The ratios of \neiiia/\neiiib$=3.320$ and \oiiib/\oiiia$=2.983$ (which are independent of ISM conditions) are derived using \pyneb \citep{luri15}. For each Balmer line (\hd, \hg, and \hb), we include both narrow and broad components. All lines are modelled as Gaussians with the same redshift, while the intrinsic linewidth ($FWHM_{\rm N}$) of all narrow lines are set to be identical. The intrinsic FWHM of the broad component of the Balmer lines are separately set to be identical ($FWHM_{\rm B, Ba}$).

It is important to note that due to flux and wavelength calibration differences, the spectra extracted from JWST/NIRSpec prism and grating data cubes spectra may not yet be compared on a 1:1 basis. Works investigating JWST/NIRSpec MSA spectra have found that the medium-resolution ($R\sim1000$) grating flux is $\sim10-15\%$ higher than the prism flux (but with considerable scatter; e.g., \citealt{bunk24,deug25_DR3}). The same works also find a redshift difference of $z_{\rm R100}-z_{\rm R1000}\simeq0.004$, which corresponds to a velocity offset of $\sim125$\,km\,s$^{-1}$ at $z=8.5$. Works using both prism and high-resolution ($R\sim2700$) data have found an even higher redshift difference ($z_{\rm R100}-z_{\rm R2700}\simeq0.006-0.009$; \citealt{pere24,jone25b}; $\sim190-285$\,km\,s$^{-1}$ at $z=8.5$). The ratio of line fluxes measured in R100 data to those measured in R2700 data has been found to vary widely between NIRSpec IFU observations ($F_{\rm R100}/F_{\rm R2700}\sim0.35-1.70$; \citealt{arri24,ji24,ji25,scho25}). With these results in mind, we allow the systemic redshift to vary between the R100 and R2700 data, and include a wavelength-independent flux ratio between the two observed spectra.

Similarly to Section \ref{bluefitsec}, we use LMFIT in `least\_squares' mode to fit our model to the data. But in this case, we fit the model to both the R100 and R2700 data at the same time. The free parameters are: the continuum slope ($\alpha_{\rm opt}$) and normalisation at $\lambda_{\rm rest}=4959\angstrom$ ($C_{\rm opt}$), the systemic redshift in each dataset ($z_{\rm R100}$, $z_{\rm R2700}$), the flux ratio of the R100 and R2700 data ($F_{\rm R100}/F_{\rm R2700}$), the intrinsic linewidths of the narrow and broad components ($FWHM_{\rm N}$, $FWHM_{\rm B, Ba}$), and the fluxes of each line (\oiiab, \neiiiab, \oiiic, \oiiiab, and the narrow and broad components of \hd, \hg, and \hb). These properties are used to create model R100 and R2700 spectra that account for the LSF, spectral gridding, flux offset, and velocity offset of each dataset. To perform the simultaneous fit, we calculate and concatenate the residual spectrum (i.e., difference between data and model, weighted by the inverse error of each dataset) for each observed spectrum and use LMFIT to find the parameters that minimize this residual.

The flux ratio of \oiia/\oiib is a strong tracer of electron density (e.g., \citealt{isob23}), varying from 0.67-2.88 for $\log_{10}(n_{\rm e}/{\rm cm}^{-3})=0-8$ (as derived with \pyneb, assuming $T_{\rm e}=10^4$\,K). While higher temperatures result in a higher ratio, most of the variation is due to density (e.g., \oiia/\oiib=0.69-2.96 for the same density range, but assuming $T_{\rm e}=10^7$\,K). The R100 data does not resolve the doublet, and the R2700 data does not include a strong detection of both lines. Because of this, it is not possible to place a strong constraint on the \oiia/\oiib ratio as a free parameter. Instead, we tested fits using a range of physically motivated \oiia/\oiib ratios ($0.7,0.8,\ldots,3.0$) and recorded the goodness of fit values for each run (i.e., $\chi^2$, reduced $\chi^2$, and Bayesian Information Criterion [BIC]). The best fit was found to be \oiia/\oiib$=0.7$, with increasingly poor fits for higher ratios. Thus, we fix \oiia/\oiib$=0.7$ in our model. This corresponds to an estimated density  of $\log_{10}(n_{\rm e}/\rm cm^{-3})\sim1.7$ for $T_{\rm e}\sim10^4$\,K, or $\log_{10}(n_{\rm e}/\rm cm^{-3})\sim2.4$ for more extreme temperatures ($T_{\rm e}\sim10^7$\,K).

We note that while an initial fit was performed with broad and narrow components for each Balmer line (\hb, \hg and \hd), this resulted in a poor fit. Because \hg falls into the detector gap for the R2700 data, we cannot measure the narrow component of this line. The fit was improved by including broad components for all three lines, and a narrow component for \hb. The narrow component of \hb is found to have a flux$\sim1/6^{\rm th}$ that of the broad component, so our current data do not allow us to detect the narrow components of the weaker \hd or \hg. 

\subsubsection{Fitting results}

\begin{figure*}
    \centering
    \includegraphics[width=\textwidth]{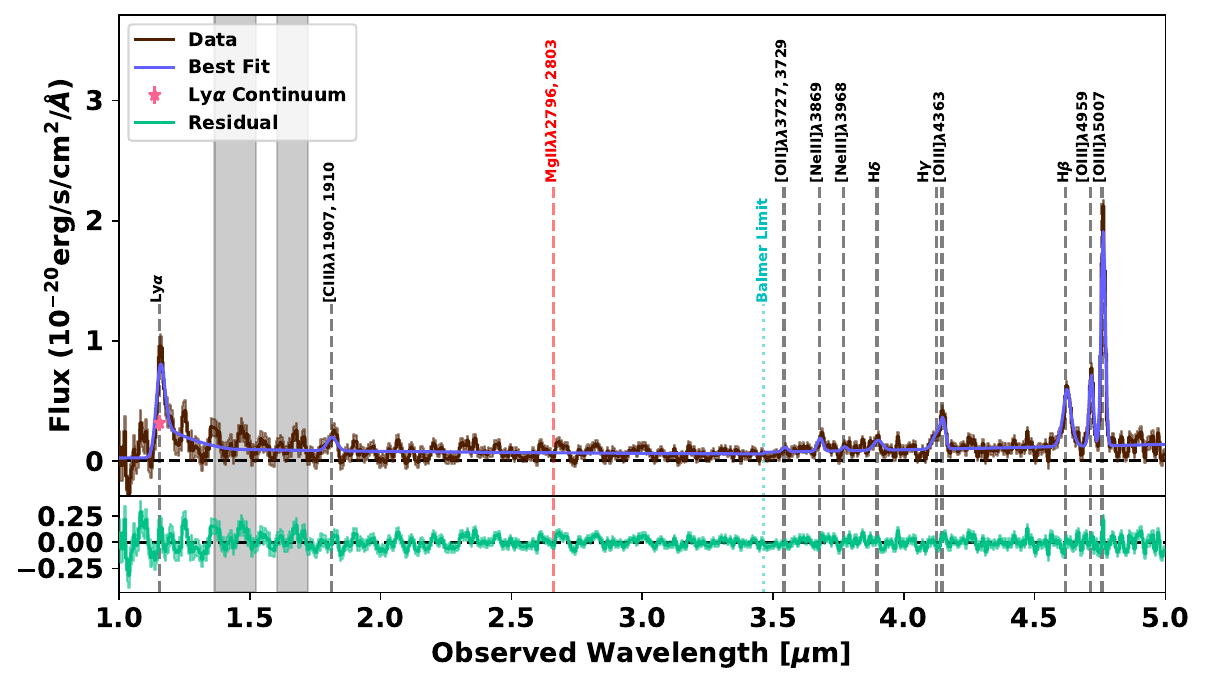}
    \caption{Spectrum extracted from NIRSpec IFU R100 data cube (brown line) using a circular aperture of radius $0.125''$. A concatenation of the best-fit red model (using R100 data only) and blue model (using R100 and R2700 data) is shown by the blue line, where the dividing wavelength (i.e., the Balmer break) is shown by a dotted cyan line. Model residuals are shown in the lower portion of the panel (green line). The best-fit centroid wavelengths of the narrow components of each line are shown by dashed vertical black lines, whereas red dashed lines indicate undetected emission lines. The best-fit continuum underlying \lya is shown by a pink star. Uncertainties ($1\sigma$) for the extracted spectrum and residuals are shown by shaded regions. The wavelength ranges that were excluded from the fit to reduce contamination by faint rest-UV lines are shown by grey shaded regions.}
    \label{gfit100}
\end{figure*}

\begin{figure*}
    \centering
    \includegraphics[width=\textwidth]{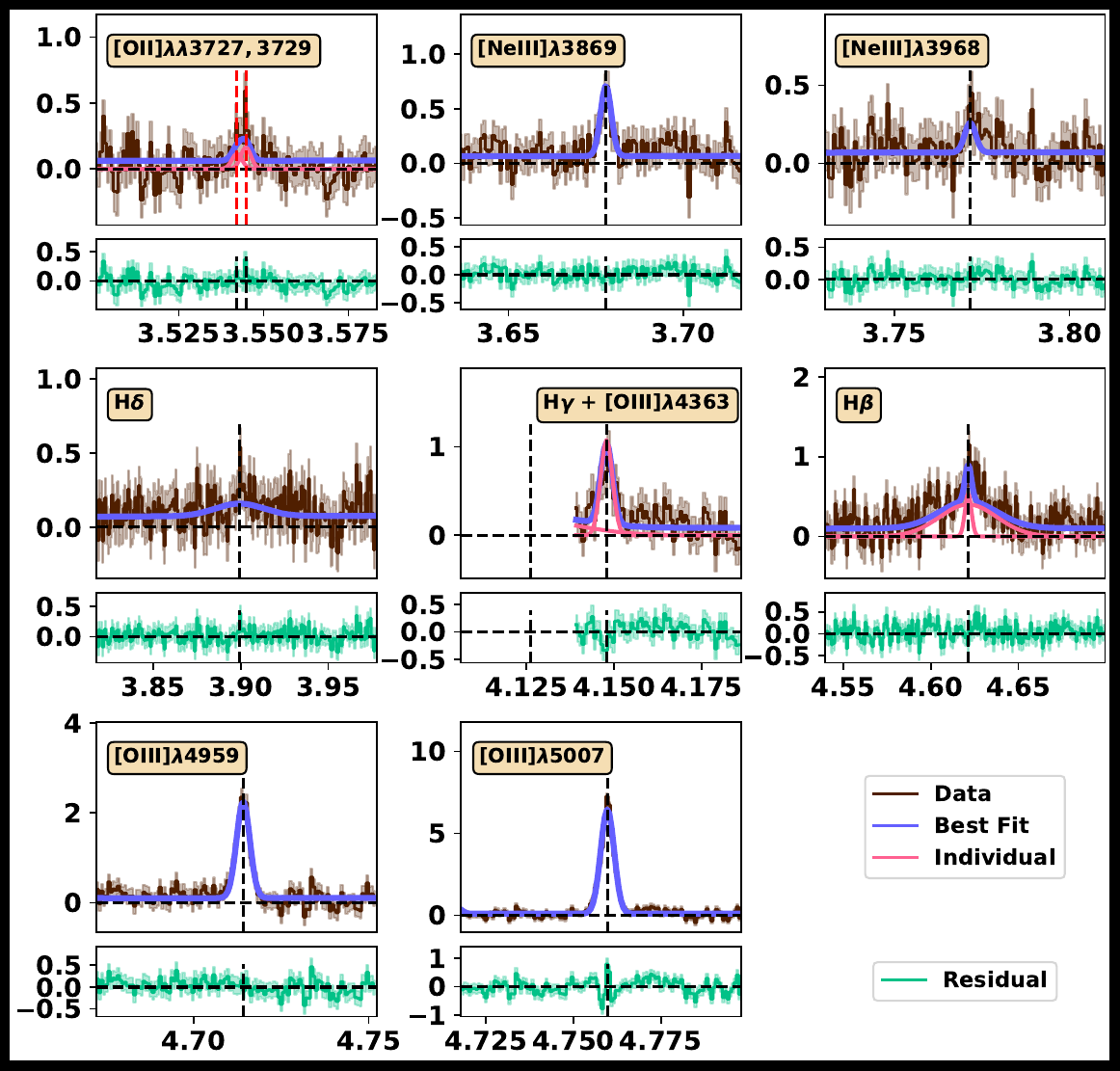}
    \caption{Spectrum extracted from NIRSpec IFU R2700 data cube (brown line) using a circular aperture of radius $0.125''$. The best-fit model (which was fit to the R100 and R2700 data simultaneously) is shown as blue lines. When two components or lines are overlapping, they are shown by pink lines. Model residuals are shown in the lower portion of each panel (green lines). The best-fit centroid wavelengths of the narrow components of each line are shown by dashed vertical lines. We only show the wavelength range around emission lines of interest. Uncertainties ($1\sigma$) for the extracted spectrum and residuals are shown by shaded regions.}
    \label{gfit2700}
\end{figure*}

Following the fitting procedure of Sections \ref{bluefitsec} and \ref{redfitsec}, we derive best-fit continuum and line parameters for \UC from both the R100 and R2700 data. The resulting R100 and R2700 fits are shown in Figures \ref{gfit100} and \ref{gfit2700}, respectively. The best-fit blue and red models for the R100 data are concatenated to create a complete model. The best-fit values are listed in Table \ref{linefluxtable}. 

Both spectra are well fit by the same intrinsic model, but we find a significant velocity offset ($182\pm10$\,km\,s$^{-1}$) and flux ratio ($0.85\pm0.01$) between the R100 and R2700 data. As noted in Section \ref{redfitsec}, these values are expected from calibration differences. The flux ratio is identical to the value assumed by \citet{deug26}, and falls within the scatter of $F_{\rm R100}/F_{\rm R2700}$ ratios found by \citet{scho25}. Other works find flux ratios of $>1$ (e.g., \citealt{arri24,ji24}), which may be due to calibration differences (e.g., updates to calibration files) or an intrinsic scatter in this ratio. The velocity offset corresponds to $z_{\rm R100}-z_{\rm R2700}\simeq0.006$ (i.e., larger than the uncertainty of the R100-based fit of \citealt{koko23}; $\delta z=0.003$), while the flux ratio implies that the grating flux is $\sim17\%$ higher than the prism flux. Each value is comparable to previous results (e.g., \citealt{bunk24,deug25_DR3,pere24,jone25b}). For our analysis, we adopt the redshift and fluxes in the R100 frame.

The residuals in each fit are generally low (i.e., $<2\sigma$).  One exception is the fit of \oiiib in the R2700 data (Figure \ref{gfit2700}), which features a central flux excess with neighbouring negative troughs. This residual pattern has been interpreted to indicate the need for a second Gaussian component, possibly representing an outflow (e.g., \citealt{gino20}). This hypothesis is weakened by the fact that the addition of an outflow component to the model did not result in a better fit. Alternatively, it may be an example of the wavelength-dependent continuum oscillations (or `wiggles') seen in NIRSpec IFU data around bright emission lines (e.g., \citealt{pern23}). However, these wiggles are artifacts caused by single-spaxel extraction, and extend over wide swathes of each spectrum. Since we use a circular aperture, and the features are isolated to the proximity of \oiiib, these are likely not wiggles. An additional possibility is that these artifacts are caused by a non-Gaussian LSF. In any case, the low amplitude of the residuals allows us to neglect them in our analysis.

In Table \ref{linefluxtable}, we include a comparison to the values of  \citet{koko23}, who observed \UC using the JWST/NIRSpec MSA in R100. We find that the best-fit linewidths, redshift, and most emission line fluxes are in agreement (i.e., within $3\sigma$). Some lines show different flux (as explored in Appendix \ref{ifumsa}), but
we find a similar placement of \UC on the line ratio diagnostic plots of \citet{mazz24}, adding credence to the classification of this source as an AGN (Section \ref{mazzsec}).

If we fit the R100 data only, then we measure a $FWHM_{\rm B,Ba}$ value within $1\sigma$ of \citet{koko23} and other works that analysed the R100 data ($\sim3500$\,km\,s$^{-1}$; \citealt{tref24,gree24}). If we fit the R2700 data alone or fit both datasets simultaneously, then a smaller $FWHM_{\rm B,Ba}\sim2500$\,km\,s$^{-1}$ is preferred (i.e., lower $\chi^2$, reduced $\chi^2$, and Bayesian Inference Criterion). If we instead fix $FWHM_{\rm B,Ba}$ to the larger value, then the R100 fit is not significantly affected, while the residuals around the \hb line in the R2700 fit are increased. This difference may be affected by the much finer fiducial spectral resolution of the R2700 data at the redshifted wavelength of \hb ($FWHM_{\rm v}\sim100$\,km\,s$^{-1}$) \textit{vs.} that of the R100 data ($FWHM_{\rm v}\sim1250$\,km\,s$^{-1}$; where the actual resolution of each dataset could be finer by a factor of $\lesssim1.7$; e.g., \citealt{koko23,degr24}). Alternatively, the LSF could deviate from a perfect Gaussian (which will be investigated in future works). But since the $FWHM_{\rm B,Ba}$ values agree to within $3\sigma$, the R100 data is well fit by either value, and the R2700 data (which features higher spectral resolution) is better fit by the narrower value, we will use this value in our analysis.

\begin{table*}
    \centering
    \begin{tabular}{c|ccccc}
  Quantity									& R100 & R2700 & Units												  &	This work									& \citet{koko23} \\ \hline
	$z_{\rm sys}$								& X	   & X	   & -													  & $8.5095\pm0.0003$ & $8.502\pm0.003$ \\
	$\Delta v_{\rm R100-R2700}$				& X	   & X	   & [km s$^{-1}$]										  & $182\pm10$ & $-$	\\
	$FWHM_{\rm N}$								& X	   & X	   & [km s$^{-1}$]										  & $270\pm5$ & $203\pm154$	\\
	$FWHM_{\rm B, Ba}$							& X	   & X	   & [km s$^{-1}$]										  & $2503\pm176$ & $3439\pm413$	\\ \hline
	$\beta_{\rm Ly\alpha}$					& X	   &	   & -													  & $-5.7\pm0.9$ & $-$	\\
	$\beta_{\rm UV}$							& X	   &	   & -													  & $-0.6\pm0.1$ & $-$	\\
	$C_{1500}$									& X	   &	   & [$10^{-20}$ erg s$^{-1}$ cm$^{-2}$ $\angstrom^{-1}$] & $0.10\pm0.01$ & $-$	\\
	$\alpha_{\rm opt}$						& X	   & X	   & -													  & $1.8\pm0.2$ & $-$	\\
	$C_{4959}$									& X	   & X	   & [$10^{-20}$ erg s$^{-1}$ cm$^{-2}$ $\angstrom^{-1}$] & $0.12\pm0.01$ & $-$	\\ \hline
	$F_{\rm Ly\alpha}$						& X	   &	   & [$10^{-20}$ erg s$^{-1}$ cm$^{-2}$]                  & $256.0\pm23.2$ & $414.1\pm85.2$	\\
	$REW_{\rm Ly\alpha}$	                    & X	   &       & [$\angstrom$]             					      & $85\pm19$ & $240\pm30$	\\ \hline
	$F_{\rm [CIII]\lambda\lambda1907,1910}$  & X	   &       & [$10^{-20}$ erg s$^{-1}$ cm$^{-2}$]                  & $52.1\pm12.1$ & $-$	\\
	$F_{\rm MgII\lambda\lambda2796,2803}$	& X	   &	   & [$10^{-20}$ erg s$^{-1}$ cm$^{-2}$]                  & $<34.8$ & $62.5\pm15.1$	\\
	$F_{\rm [OII]\lambda\lambda3727,3729}$	& X	   & X	   & [$10^{-20}$ erg s$^{-1}$ cm$^{-2}$]                  & $10.4\pm2.0$ & $15.0\pm8.0$	\\
	$F_{\rm [NeIII]\lambda3869}$	            & X	   & X	   & [$10^{-20}$ erg s$^{-1}$ cm$^{-2}$]                  & $29.3\pm2.5$ & $62.2\pm15.4$ \\
	$F_{\rm H\delta,N}$						& X	   & X	   & [$10^{-20}$ erg s$^{-1}$ cm$^{-2}$]                  & $<23.3$ & $12.2\pm6.5$	\\
	$F_{\rm H\delta,B}$						& X	   & X	   & [$10^{-20}$ erg s$^{-1}$ cm$^{-2}$]                  & $34.9\pm5.0$ & $-$	\\
	$F_{\rm H\gamma,N}$						& X	   & X	   & [$10^{-20}$ erg s$^{-1}$ cm$^{-2}$]                  & $<27.2$ & $-$	\\
	$F_{\rm H\gamma,B}$						& X	   & X	   & [$10^{-20}$ erg s$^{-1}$ cm$^{-2}$]                  & $59.2\pm7.9$ & $-$	\\
	$F_{\rm [OIII]\lambda 4363}$				& X	   & X	   & [$10^{-20}$ erg s$^{-1}$ cm$^{-2}$]                  & $45.9\pm3.6$ & $83.0\pm7.2$	\\
	$F_{\rm H\beta,N}$						& X	   & X	   & [$10^{-20}$ erg s$^{-1}$ cm$^{-2}$]                  & $25.0\pm5.0$ & $78.6\pm5.9$	\\
	$F_{\rm H\beta,B}$						& X	   & X	   & [$10^{-20}$ erg s$^{-1}$ cm$^{-2}$]                  & $164.8\pm9.4$ & $232.6\pm17.3$	\\
	$F_{\rm [OIII]\lambda 5007}$			    & X	   & X	   & [$10^{-20}$ erg s$^{-1}$ cm$^{-2}$]                  & $359.6\pm5.1$ & $412.6\pm11.3$	\\ \hline
	$F_{\rm R100}/F_{\rm R2700}$				& X	   & X	   & -                 									  &$0.85\pm0.01$ & $-$	\\ \hline
\end{tabular}
    \caption{Best-fit emission line properties, as derived from fits to R100 and R2700 data. The two datasets feature a significant velocity offset ($\Delta v_{\rm R100-R2700}$) and flux ratio ($F_{\rm R100}/F_{\rm R2700}$), which are listed here. Redshift and fluxes are reported as measured in the R100 frame. We mark which properties were derived with the R100 data (second column) and the R2700 data (third column). All values have been corrected for gravitational magnification and aperture losses. For undetected lines, we present $3\sigma$ upper limits on their flux. For \lya, we present both the best-fit flux ($F_{\rm Ly\alpha}$) and rest-frame equivalent width ($EW_{\rm 0,Ly\alpha}$).}
    \label{linefluxtable}
\end{table*}

\section{Properties of \UC}\label{analysis_sec}

The spectral fits of the previous Section result in a number of constraints on line and continuum properties of \UC. Here, we utilise these best-fit values to characterise the black hole and host galaxy of this system.

\subsection{Dust attenuation}\label{dasec}

Previously, \citet{koko23} measured a best-fit ratio of Balmer line fluxes ($(F_{\rm H\gamma}/F_{\rm H\beta})_{\rm obs}=0.31\pm0.06$) and derived $A_V$ using:
\begin{equation}
A_{\rm V} = R_{\rm V} \frac{2.5}{k(\lambda_{\rm H\beta})-k(\lambda_{\rm H\gamma})} \log_{10}\left[ \frac{(F_{\rm H\gamma}/F_{\rm H\beta})_{\rm obs}}{(F_{\rm H\gamma}/F_{\rm H\beta})_{\rm int}}\right]
\end{equation}
assuming case B recombination and the reddening law found for the bar of the Small Magellanic Cloud (SMC; \citealt{gord03}). This reddening law is commonly adopted for high-redshift quasars (e.g., \citealt{gree24,mazz24_CEERS,tref24,deug25,ji25}). Using this line flux ratio and an intrinsic ratio of $(F_{\rm H\gamma}/F_{\rm H\beta})_{\rm int}=1/2.14$ results in $A_{\rm V}=2.1^{+1.1}_{-1.0}$, or a colour excess of $E(B-V)\equiv A_{\rm V}/R_{\rm V}=0.8\pm0.4$. However, the \hg flux used in this calculation included a contribution from the broad component, while the \hb flux was from the narrow component only.

We derive further constraints on $E(B-V)$ using all observed Balmer decrements. We only consider pairs of lines from the same emission origin (narrow or broad). The intrinsic ratios are derived using \pyneb, assuming the ISM conditions found for a stack of $4<z<7$ Type I AGN from JADES \citep[][$T_{\rm e}\sim2\times10^4$\,K, $n_{\rm e}\sim10^4$\,cm$^{-3}$]{isob25}. As seen in Table \ref{prop_table}, we are able to estimate the colour excess of the broad component using three ratios (\hd/\hb, \hg/\hb, and \hd/\hg). All three of these ratios do not show strong evidence for significant dust attenuation, with $E(B-V)$ values within $2\sigma$ of 0. While the small wavelength difference between these lines makes measurement of attenuation more difficult, other JWST/NIRSpec IFU studies of galaxies at $z>7$ (where \ha is not covered by the nominal wavelength range of JWST/NIRSpec) have found similar results based on Balmer decrements (e.g., \citealt{bunk23,jone24_B14,zamo25b}). While we find no significant evidence for reddening in the broad or narrow Balmer lines , we note that JWST/MIRI observations of \ha would enable more precise constraints.

Of course, this result is dependent on a few assumptions. First, we adopt the SMC bar extinction curve of \citet{gord03}. Because each dust attenuation curve is normalised by the value at the approximate wavelength of the \hb-\oiiiab complex (i.e., V-band), the choice of extinction law does not introduce strong changes in the observed values of these lines. As a test, we adopt the starburst reddening law of \citet{calz00}, resulting in nearly identical values (i.e., $<1\sigma$ discrepant). However, we note that the choice of extinction curve will affect the observed rest-UV line fluxes and the continuum slope.

Next, we consider the fact that the intrinsic Balmer ratios are derived using \pyneb and our fiducial ISM conditions. For most densities ($\log_{10}(n_{\rm e}/{\rm cm}^{-3})\lesssim 6$), these ratios feature a slight dependence on density (e.g., \hg/\hb changes by $\lesssim5\%$ between $\log_{10}(n_{\rm e}/{\rm cm}^{-3})=0-7$ and $T_{\rm e}=(0.5-3.0)\times10^4$\,K). At extreme densities ($\log_{10}(n_{\rm e}/{\rm cm}^{-3})\gtrsim 7$), the ratio transitions to be density dependent. Changing the temperature and density in these ranges does not result in a significant (i.e., $>1\sigma$) change in the $E(B-V)$ value. 

An additional possibility is that we are observing a system where case B recombination is not applicable (e.g., \citealt{pirz24,scar24,mccl25,sun25}). Our observed Balmer decrements suggest minimal dust attenuation (assuming case B), but the fact that our derived $E(B-V)$ values are not significantly negative does not provide strong evidence against case B.

We note that \citet{koko23} reported a large dust attenuation ($A_{\rm V}\sim1.9$) from fitting an AGN-only model to the continuum of \UC. At face value, this appears to disagree with the lack of strong attenuation found through analysis of Balmer ratios. However, as noted by other works, AGN-only models return higher dust attenuation than star+AGN models (e.g., \citealt{leun25}). In addition, the Balmer and continuum emission may emerge from different regions, or feature a different geometry. Thus, this difference in $A_{\rm V}$ could be used as evidence for a stellar component that contributes to the continuum emission, and/or a dusty AGN surrounded by less obscured line-emitting gas.

The non-detection of warm dust emission in a sample of 60 LRDs at $5\leq z \leq 8$ \citep{case25} suggests that this class of objects may not feature significant warm dust reservoirs (i.e., $M_{\rm dust}<10^6\,M_{\odot}$). \UC only features a single ALMA non-detection, from the Deep UNCOVER-ALMA Legacy High-z (DUALZ) survey \citep{fuji23}. They do not detect significant $\lambda_{\rm obs}=1.2$\,mm emission from \UC, implying a $3\sigma$ upper limit of $S_{\rm 1.2mm}< 99\,\mu$Jy. From this single limit, it is not possible to solve the degeneracy inherent in FIR SED modelling (i.e., between dust temperature $T_{\rm dust}$, dust mass $M_{\rm dust}$, dust emissivity index $\beta_{\rm IR}$). 

Assuming a modified blackbody model that accounts for the effect of the CMB at high redshift (e.g., \citealt{carn19}) and adopting a standard value of $\beta_{\rm IR}=1.8$ suitable for most high-redshift galaxies (e.g., \citealt{wits23}), we find that this observed flux density upper limit does not constrain the dust mass (i.e., even models with $M_{\rm dust}=10^{10}$\,M$_{\odot}$ are in agreement) in the case of cold gas ($T_{\rm dust}=30$\,K, or the lower limit for LRDs found by \citealt{li25}). By adopting the conservative limit of $M_{\rm dust}\leq M_*<10^{9.7}\,M_{\odot}$, we find that the upper limit is only met for $T_{\rm dust}\lesssim40$\,K. Thus, while we may rule out a large reservoir of hot ($T_{\rm dust}>40$\,K) dust, additional ALMA observations (i.e., deeper and at different wavelengths) are required to say more. 

\begin{table}
\centering
\begin{tabular}{c|cc}
Quantity                        &   This work           & \citet{koko23}\\ \hline
$E(B-V)_{\rm H\delta/H\beta,B}$	&	$0.3\pm0.2$	&	-	\\
$E(B-V)_{\rm H\gamma/H\beta,B}$	&	$0.5\pm0.3$	&	-	\\
$E(B-V)_{\rm H\delta/H\gamma,B}$	&	$-0.1\pm0.7$	&	-	\\ \hline
$\log_{10}(M_{\rm \BH,GH05,H\beta}/M_{\odot})$	&	$7.45\pm0.40$	&	$8.17\pm0.42$	\\
$\log_{10}(M_{\rm \BH,VP06,H\beta}/M_{\odot})$	&	$7.57\pm0.40$	&	-	\\
$\log_{10}(M_{\rm \BH,VP06,5100}/M_{\odot})$	&	$7.58\pm0.40$	&	$8.01\pm0.40$	\\ \hline
$\log_{10}(L_{\rm bol}/{\rm erg s}^{-1}$)	&	$44.71\pm0.02$	&	$45.82^{+0.17}_{-0.28}$\\ \hline	
$\log_{10}(L_{\rm Edd,GH05,H\beta}/{\rm erg s}^{-1}$)	&	$45.6\pm0.4$	&	-	\\
$\log_{10}(L_{\rm Edd,VP06,H\beta}/{\rm erg s}^{-1}$)	&	$45.7\pm0.4$	&	-	\\
$\log_{10}(L_{\rm Edd,VP06,5100}/{\rm erg s}^{-1}$)	&	$45.7\pm0.4$	&	-	\\ \hline
$\lambda_{\rm Edd,GH05,H\beta}$				&	$0.15^{+0.23}_{-0.09}$				&		$\sim 0.40$	\\
$\lambda_{\rm Edd,VP06,H\beta}$				&	$0.11^{+0.17}_{-0.07}$				&		-	\\
$\lambda_{\rm Edd,VP06,5100}$					&	$0.11^{+0.17}_{-0.07}$				&		-	\\ \hline
\end{tabular}
\caption{Properties of \UC, as derived from spectral fits to R100 and R2700 data. All values have been corrected for gravitational magnification and aperture losses.}
\label{prop_table}
\end{table}

\subsection{Black hole mass estimation}\label{bhmass}

The black hole mass may be estimated using the properties of the broad \hb line, which originates from the BLR nearby the black hole, and so-called single-epoch virial relations. The most common method is that of \citet{gree05}:
\begin{multline}
\log_{10}(M_{\rm \BH,GH05,H\beta}/M_{\odot}) = 
(6.56\pm0.02) +\\
(0.56\pm0.02)\log_{10}\left( \frac{L_{\rm H\beta}}{10^{42}{\rm erg\,s^{-1}}}\right) +
2\log_{10}\left( \frac{FWHM_{\rm H\beta}}{10^{3}{\rm km\,s^{-1}}}\right)
\end{multline}

We also consider the relations of \citet{vest06}, who derived a similar method based on \hb:
 \begin{multline}
\log_{10}(M_{\rm \BH,VP06,H\beta}/M_{\odot}) = (6.67\pm0.03) \\ + 
0.63\log_{10}\left( \frac{L_{\rm H\beta}}{10^{42}{\rm erg\,s^{-1}}}\right) + 
2\log_{10}\left( \frac{FWHM_{\rm H\beta}}{10^{3}{\rm km\,s^{-1}}} \right)
\end{multline}
and a relation from the same work that replaces the \hb luminosity with the continuum emission underlying \oiiib ($\lambda_{\rm rest}=5100\angstrom$):
 \begin{multline}
\log_{10}(M_{\rm \BH,VP06,5100}/M_{\odot}) = (6.91\pm0.02) \\ + 
0.50\log_{10}\left( \frac{\lambda L_{\lambda}(5100\AA)}{10^{44}{\rm erg\,s^{-1}}}\right) + 
2\log_{10}\left( \frac{FWHM_{\rm H\beta}}{10^{3}{\rm km\,s^{-1}}} \right)
\end{multline}
where each relation has an intrinsic scatter of $\sim0.4$\,dex.

We implement these relations with the best-fit properties of the R100 and R2700 data, resulting in values of $\log_{10}(M_{\BH}/M_{\odot})=7.4-7.6$, with uncertainties of $\sim0.4$\,dex (i.e., with uncertainties dominated by the scatter of each relation; Table \ref{prop_table}). This value is lower than the value of $\log_{10}(M_{\BH}/M_{\odot})=8.0-8.2$ (with uncertainties of $\sim0.4$\,dex) derived by \citet{koko23}, but not significantly ($<2\sigma$ discrepant). This different value is partially due to the fact that we do not detect significant evidence for dust attenuation from the Balmer decrement, while \citet{koko23} corrected \hb using $A_{\rm V}\sim2$. In addition, the other work found a larger FWHM and line flux of the broad component of \hb. 

It is worthwhile noting that the single-epoch methods have been calibrated on galaxies in the local Universe, and their application to high-redshift sources is not straightforward. 
Indeed, recent works have suggested that, for LRDs, the broad emission is due to electron scattering rather than bulk motion around a black hole (\citealt{rusa25}), resulting in a much lower ($\geq2$\,dex) $M_{\rm BH}$ estimate. As a test, we fit our spectra using the electron scattering assumption, resulting in a slightly better fit (i.e., $\Delta \rm BIC\sim3$). While this would move many high-redshift AGN onto the local $M_{\rm BH}-M_*$ relation (see Section \ref{disc_sec}), this interpretation is the subject of debate (e.g., \citealt{bosm25,braz25,juod25}). Thus, we adopt the currently common bulk motion assumption for our analyses.
More recently, \citet{juod25_LRD} directly measured the black hole mass in a lensed LRD at $z=7$ by resolving its sphere of influence, finding a mass fully consistent with the virial relations. We consider further these results for our discussion of the black hole {\it vs.\ }galaxy properties in Section \ref{disc_sec}.

\subsection{Luminosities \& Eddington ratio}

The bolometric luminosity can be found from the \hb broad luminosity using the relation from \citet{ster12}:
\begin{equation}
L_{\rm bol} = 130\times L_{\rm H\beta,B}\times(F_{\rm H\alpha}/F_{\rm H\beta}) 
\end{equation}
where $(F_{\rm H\alpha}/F_{\rm H\beta})=2.74$ is the intrinsic ratio of these lines, assuming case B recombination and the fiducial ISM conditions of Section \ref{dasec}\footnote{Assuming a higher density ($\log_{10}(n_{\rm e}/{\rm cm}^{-3})=7$), as suggested by the analysis Section \ref{mazzsec}, results in a fiducial $(F_{\rm H\alpha}/F_{\rm H\beta})=2.71$, or a decrease of $\sim1\%$.}.

We find a value of 
$\log_{10}(L_{\rm bol})=44.72\pm0.02$,  which is considerably lower than the value of 
$\log_{10}(L_{\rm bol})=45.82^{+0.17}_{-0.28}$ found by \citet{koko23}. This difference is partially ascribable to our different treatments of dust attenuation (see Section \ref{dasec}). If a high attenuation of $A_{\rm V}=2$ is taken into account, then the intrinsic \hb flux increases by a factor of $\sim20\times$, resulting in higher estimates of $L_{\rm bol}$. Similarly, the higher uncertainty in the previous work is due to the uncertain dust correction, while our value contains no dust correction. As a test, we applied this dust correction to our values, resulting in $L_{\rm bol}$ values within $1\sigma$ of \citet{koko23}.

Similarly, the Eddington luminosity (e.g., \citealt{fabi12}) is expressed as:
\begin{equation}
L_{\rm Edd} = \frac{4\pi G M_{\BH}m_{\rm p}c}{\sigma_{\rm T}}
= 1.26\times10^{38} \left( \frac{M_{\BH}}{M_{\odot}}\right)[\rm erg\,s^{-1}]
\end{equation}
This is commonly used to calculate the Eddington ratios ($\lambda_{\rm Edd}\equiv L_{\rm bol}/L_{\rm Edd}$). Using the \citet{gree05} method to derive black hole mass, we find $\lambda_{\rm Edd}=0.15^{+0.23}_{-0.09}$ (which is in agreement with the ratios derived using other black hole mass estimation methods). Our ratio is lower than the value of $\sim0.40$ measured by \citet{koko23}, but due to the large uncertainty on $L_{\rm Edd}$, they are $<2\sigma$ discrepant. We note that both $L_{\rm bol}$ and $L_{\rm Edd}$ are functions of $L_{\rm H\beta}$ (using the calibration of \citealt{gree05}), and a higher \hb flux (or a lower $FWHM$ of the broad component of \hb) will result in a higher $\lambda_{\rm Edd}$. Since we adopt the same equations, this lower ratio is due to a combination of different dust attenuation assumptions and different aperture loss corrections, as well as the different broad \hb flux and smaller \hb broad linewidth.

\subsection{ISM conditions}\label{ism_sec}

While the broad emission opens a window into the nature of the black hole of \UC, the narrow lines instead allow us to characterise the host galaxy (or nearby gas; e.g., \citealt{inay25,lin25_env}). 

To begin, we use our observed line ratios and the code \pyneb to explore the electron density ($n_{\rm e}$) through the \oiiab line pair. As noted in Section \ref{redfitsec}, we are not able to directly fit for the ratio of \oiia/\oiib using our data. An exploration of line ratios finds that a low ratio ($\sim0.7$, implying $\log_{10}(n_{\rm e}/{\rm cm}^{-3})\sim 2$) yields a better fit. While this ratio has a weak dependence on $T_{\rm e}$, a low ratio may only be caused by a low-density environment (i.e., at any temperature).


Next, we consider the \oiiic/\oiiib ratio, which is a strong indicator of $T_{\rm e}$. This ratio is lower in our data ($0.13\pm0.01$) than that of \citet[][$0.20\pm0.02$]{koko23}. As noted in \citet{koko23}, this ratio is much higher than expected for most galaxies. For an electron density of $n_{\rm e}\sim10^{1-4}$\,cm$^{-3}$, \pyneb predicts a maximum ratio of $\lesssim0.05$ for $T_{\rm e}<2.5\times10^4$\,K. While this line ratio is strongly dependent on $T_{\rm e}$ for $n_{\rm e}<10^5\,{\rm cm}^{-3}$, it transitions to a strong dependence on $n_{\rm e}$ for $n_{\rm e}>10^5\,{\rm cm}^{-3}$. While a ratio of $0.1-0.2$ could be theoretically explained by a high density ($n_{\rm e}\sim10^6$\,cm$^{-3}$) and temperature ($\sim2\times10^4$\,K), this density is $\sim2$\,dex higher than the average value found for $4<z<7$ Type I AGN \citep{isob25} and $\sim3$\,dex higher than the average value for star-forming galaxies (SFGs) at the redshift of \UC (e.g., \citealt{isob23,li25ne}). While high $n_{\rm e}$ is expected from some interpretations of LRD properties (e.g., \citealt{rusa25,bege26}), these conditions are extreme. 

Together, these tracers appear to present contrasting ISM conditions: the low \oiia/\oiib ratio requires low $n_{\rm e}$, while the high \oiiic/\oiiib ratio requires high $n_{\rm e}$. One possible explanation is a two-phase ISM, with both low- and high-density emission regions (e.g., \citealt{ji24,hari25,usui25}). Alternatively, the \oiiic/\oiiib ratio could suggest an extraordinarily high temperature (i.e., $>2\times10^4$\,K). Similarly high \oiiic/\oiiib ratios have been observed in other $z>7$ galaxies (e.g., \citealt{scha22,katz23,cull25}), where they have been explained by a high degree of cosmic ray heating, the presence of high-mass X-ray binaries, or Wolf-Rayet stars. The extreme conditions implied by these ratios independently support the presence of an AGN in this galaxy (see \citealt{mazz24,uble24}, Section \ref{mazzsec}).
These aspects will be discussed more thoroughly in the next Section, within the context of AGN and SF photoionization models.


Our observed line fluxes can also be used to estimate the gas-phase metallicity of \UC using strong line ratio diagnostics (e.g., \citealt{curt20,cata25}). First, we use the calibrations of \citet{sand24} for SFGs, as they were created using JWST observations of galaxies in a redshift range that contains \UC (i.e., $2.1<z<8.7$). Using the narrow-line flux ratios (O3, O2, R23, O32, Ne3O2, see Table \ref{linerat} for ratio definitions), we find a best-fit value of $12+\log_{10}(O/H)=7.78\pm0.06$, or $0.12\pm0.02$ solar. As an alternative, we consider the rest-UV emission line metallicity diagnostics for AGN galaxies. If we adopt the calibration of \citet{dors21}, which was derived using observations of local Seyfert galaxies, we find a higher metallicity ($12+\log_{10}(O/H)=8.53\pm0.12$, or $0.69\pm0.19$ solar). The AGN models of \citet{zhu24} instead suggest a $12+\log_{10}(O/H)=8.37\pm0.10$, or $0.48\pm0.11$\,solar.

There are caveats with these metallicity derivations, including the possibility that \oiiab and \oiiiab (which are in every calibration) originate from different components of a multi-phase ISM, the possibility that the electron temperature in this object is exceptionally high, or the unproven applicability of metallicity calibration created using SFGs, pure AGN, or local Seyferts to LRDs (e.g., \citealt{mazz24}). 

Most works investigating LRDs assume a low metallicity ($\lesssim 20\%$ solar; e.g., \citealt{donn25,koce25}), in agreement with the finding of $\lesssim10\%$ solar metallicity from individual studies of LRDs (e.g., \citealt{degr25,deug25,maio25_QSO1}). Our SFG-based metallicity estimate is therefore in line with previous findings for high-$z$ LRDs, while the AGN-based calibrations return much higher metallicities. This difference may be caused by a combination of emission from a central AGN and host galaxy (e.g., \citealt{degr25_new}), although this is still under investigation.

\begin{table}
\centering
\begin{tabular}{c|c}
Line ratio & Definition \\ \hline
O3 & $F_{\rm [OII]\lambda\lambda3727,3729}/F_{\rm H\beta}$ \\
O2 & $F_{\rm [OIII]\lambda5007}/F_{\rm H\beta}$ \\
R23 & $(F_{\rm [OII]\lambda\lambda3727,3729}+F_{\rm [OIII]\lambda\lambda4959,5007})/F_{\rm H\beta}$\\
O32 & $F_{\rm [OIII]\lambda5007}/F_{\rm [OII]\lambda\lambda3727,3729}$\\
Ne3O2 & $F_{\rm [NeIII]\lambda3869}/F_{\rm [OII]\lambda\lambda3727,3729}$\\
O3Hg & $F_{\rm[OIII]\lambda4363}/F_{\rm H\gamma}$\\
O33  & $F_{\rm[OIII]\lambda5007}/F_{\rm[OIII]\lambda4363}$
\end{tabular}
\caption{Definitions of line ratios used in this work.}
\label{linerat}
\end{table}

If \oiiib and \oiiab originate from the same region, we may use our best-fit line fluxes to estimate a line ratio of $\rm O32=34\pm9$. This O32 value is larger than the average value found for a sample of $z>7$ galaxies by \citet[][$\sim12$]{tang23}, and is comparable to the most extreme O32 observed in the $z\sim4-8$ sample of \citet[][$\sim35$]{masc23} and the $z\sim5.5-9.5$ sample of \citet[][$\sim30$]{came23}. Our O32 value may then be converted to an approximate ionization parameter $U$ using the photoionization models of \citet{mori16}, resulting in an approximate value of $\log_{10}(U)=-1.3\pm0.3$. This is higher than the standard assumption of $\log_{10}(U)=-2.0$ (e.g., \citealt{chou25}), and approaches the theoretical limit of $\log_{10}(U)<-1.0$ for HII regions (e.g., \citealt{yeh12}). We emphasize that there are ambiguities concerning these values (O32 and $U$) due to poor detection of \oiia and \oiib, a possible multi-phase ISM, and the use of a photoionization model that may not be appropriate for a high-redshift LRD. Despite this, the fact that \oiiib is well detected while \oiiab is not suggests a very high O32 in at least one component of the ISM, and thus a high level of ionisation.

\subsection{AGN diagnostics and extreme densities}\label{mazzsec}

\begin{figure*}
    \centering
    \includegraphics[width=\textwidth]{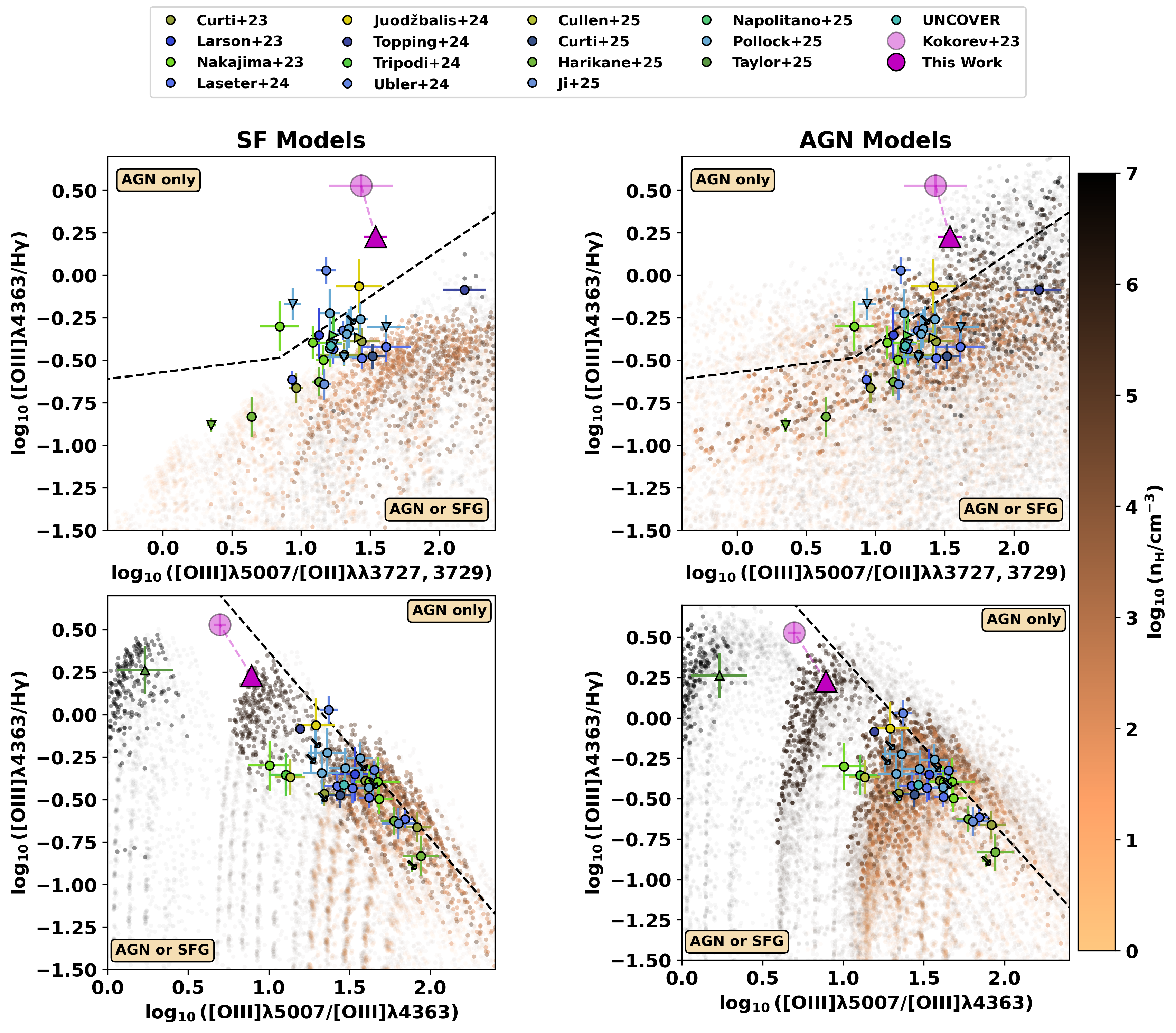}
    \caption{Line ratio diagnostic diagrams of \citet{mazz24}, with the value for \UC from \citet[][magenta circle]{koko23} and the limit from this work (magenta triangle). 
    Points above each line suggest an AGN-only nature, while those below could be interpreted as AGN or star-forming galaxies (SFGs). 
    For comparison, we include a collection of $z>6$ measurements (\citealt{curt23_mirko,lars23,naka23,lase24,juod24,topp24,trip24,uble24,cull25,curt25,hari25,ji25gnz11,napo25,poll25,tayl25}). Values from the HOMERUN models \citep{marc24_HR} are coloured by hydrogen density, with star-forming models in the left column and AGN models in the right column. Models where the density, metallicity, or ionisation parameter are comparable to our derived values are shown by darker points, while the fainter points show the full model grid.}
    \label{mazzfig}
\end{figure*}

Since our NIRSpec data do not cover the wavelength range of \ha, \niiab, or \siiab, we are unable to investigate the position of \UC on the commonly used [NII]-BPT (\citealt{bald81}) or [SII]-VO87 (\citealt{veil87}) line ratio diagrams. However, a powerful set of alternative diagnostics are those of \citet{mazz24}. In addition to O32 and Ne3O2, these diagnostics make use of the \oiiic auroral line through $\rm O3Hg=[OIII]\lambda 4363/H\gamma$ and $\rm O3= [OIII]\lambda 4363/[OIII]\lambda 5007$ (Figure \ref{mazzfig}). These are used to create line ratio diagrams, with regions in which only AGN are found, while other parts of the diagrams are populated by both AGN and star forming galaxies.

Previously, \citet{mazz24} used the measured values of \citet{koko23} to show that \UC was indeed an AGN. Our newly measured line fluxes confirm this (Figure \ref{mazzfig}). While the \hg flux reported by \citet{koko23} include contributions from both the narrow and broad components, we only use the $3\sigma$ upper limit on the narrow component of \hg (Table \ref{linefluxtable}{\footnote{Note that since \hg falls in a spectral gap of the R2700 data, this upper limit is dependent on the R100 data.}}) to find a similar result. Our limit lies below the demarcation line for the $\rm O3Hg-O33$ plot (lower panels of Figure \ref{mazzfig}), but does not disagree with an AGN interpretation, given that this lower region is populated both by AGN and SF galaxies.

To place this in a broader context, we compare the location of \UC in these line ratio diagrams to those of high-redshift ($z>6$) sources from literature (\citealt{curt23_mirko,lars23,naka23,lase24,juod24,topp24,uble24,cull25,curt25,hari25,ji25gnz11,poll25,tayl25}). We also collect all sources from UNCOVER DR4 (\citealt{beza24,pric25})\footnote{\url{https://jwst-uncover.github.io/DR4.html}} with significant line detections, resulting in one additional source (UNCOVER\_10646, $z=8.5$). From this comparison, it is clear that \UC features one of the highest O3Hg ratios among the observed galaxies. A rare exception in this sample is the upper limit of CAPERS-LRD-z9 \citep{tayl25}, which is comparable to ours. This source also features a very low O33, implying a high \oiiic/\oiiib ratio of $\sim0.6$. As discussed in Section \ref{ism_sec}, even our estimated  \oiiic/\oiiib ratio of $\sim0.1$ is difficult to explain without requiring extreme conditions (i.e., unphysically high temperature or density). While CAPERS-LRD-z9 is a BL LRD with a similar $M_{\rm BH}$ (see Section \ref{cosmictime}), it lies far from the AGN demarcation line in the $\rm O3Hg-O33$ diagnostic plot. Thus, \UC remains one of the most clear AGN in these diagnostics.

The faint coloured points in each panel represent line ratios from a large library of single-cloud CLOUDY (\citealt{ferl93,ferl17}) models used by \citet{marc24_HR} and \citet{ceci25} for the `Highly Optimized Multicloud Emission-line Ratios Using photo-ionizatioN' (HOMERUN) model. 
The star-forming galaxy CLOUDY models (left column) were computed with the following parameters. The ionizing continua are from stellar population models from BPASS v2.3 (\citealt{stan18,byrn22}) including binary stellar evolution and use a \citet{krou01} initial mass function with an upper mass cut-off of $300 \,M_{\odot}$. The modelled stellar populations featured metallicities $\log_{10}(Z_{\star})=-1.7 (solar), -2.7, -4.0, -5.0$ and ages in $\log_{10}({\rm age/Myr}) = 6.0, 6.4, 6.6, 7.1, 8.0$. The ionized gas has ionization parameter $\log_{10}(U)$ ranging from -4.0 to -0.5 in steps of 0.5, constant gas density $\log_{10}(N_{\rm H} / {\rm cm}^{-3})$ ranging from 0 to 7 in steps of 1, and metallicities $\log_{10}(Z_{\rm gas}/Z_{\odot})$ ranging from $-4.3$ to $0.4$ in steps of 0.2 (with the constraint that the gas metallicity is within a factor 0.1 and 100 of the stellar metallicity). Models were considered both with and without dust.

The ionizing continuum for the AGN models (right column) is described by a power law with UV slope $\alpha_{\rm UV} =-0.5$, an exponential cut-off $exp(-h\nu/k T_{\rm max})$ and an X-ray power-law with  slope of  $\alpha_{\rm UV} =-1.0$ linked to the UV through the $\alpha_{\rm ox}$ parameter. Models were computed with combinations of $\log_{10}(T_{\rm max}/{\rm K})= 4.0, 4.5, 5.0, 5.5, 6.0, 6.5, 7.0$ and $\alpha_{\rm ox} =-1.2,-1.5,-1.8$. The other parameters are the same with the exception that the ionization parameter upper limit is 0.5 and that the gas metallicity always cover the range of $-2\leq Z/Z_{\odot}\leq1.0$ with steps of 0.2.

The full grid of CLOUDY models spans a wide range of physical parameters, many of which unlikely to be applicable to \UC. For clarity, we isolate a subset of the models ($2<\log_{10}(N_{\rm H})<7$, $-1.5<\log_{10}(U)<-1.0$, $-1.0<\log_{10}(Z)<0.0$) that approximate the properties of our source as found in Section \ref{ism_sec} ($\log_{10}(n_{\rm H})\sim2-7$ from our analyses of the \oiiab doublet and \oiiic/\oiiib ratio, $\log_{10}(U)=-1.3\pm0.3$ from O32, and $\log_{10}(Z)\sim-0.2$ to $-0.9$ depending on calibration). This subset is shown in Figure \ref{mazzfig} as darker points.

The comparison of these SF (left panels) and AGN models (right) clearly indicate that these diagrams (and especially O3Hg vs O32) are extremely effective in identifying AGN, and that the dividing line between AGN-only and AGN+SF regions is likely very conservative. It is noticeable that even extremely high densities are not enough to reach high \oiiic/\hg values (at a given O32 value) when considering SF galaxies. In order to reach very high \oiiic/\hg the presence of an AGN is needed, likely because the hard AGN radiation field is very effective in heating the ISM, hence boosting the flux of the corona line.

The specific case of \UC is particularly intriguing, as it requires not only AGN photoionization but also extremely high densities, of about $\rm 10^7\,cm^{-3}$. Together with the fact that this is the one of the most distant sources in the sample shown, this may indicate that \UC is forming embedded in the very dense gas of an early protogalaxy.


Recently, \citet{lamb25} analysed JWST/NIRSpec MSA data of the $z=6.68476$ LRD THRILS\_46403 (also known as CEERS-10444; \citealt{koce25}), revealing evidence for a very high $\log_{10}(\rm O3Hg)=1.16\pm0.06$. This ratio is higher than observed in \UC, and is also $\sim0.5$\,dex higher than the most extreme HOMERUN AGN models, suggesting uncommon conditions. However, we note the strong Balmer absorption observed in this source is not included in the \hg line fit, resulting in an overestimation of the O3Hg ratio; additionally prominent FeII emission lines are observed in the spectrum, implying that the \oiiic is also likely contaminated by FeII. Even taking these aspects into account, THRILS\_46403 represents an extreme galaxy similar to \UC.

\subsection{Dynamical mass}\label{mdynsec}

To constrain the dynamical mass of the \UC host galaxy, we use the same relation as \citet{uble23}:
\begin{equation}
M_{\rm dyn} = K(n) K(q) \sigma_*^2 r_{\rm e} / G
\end{equation}
where $n$ is the S\'ersic index, $q$ is the axis ratio, and $K(n)=8.87-0.831n+0.0241n^2$ and $K(q)=\left[0.87+0.38e^{-3.71(1-q)}\right]^2$ are taken from \citet{cape06} and \citet{vand22}, respectively. To convert our integrated narrow gas emission linewidth (see $FWHM_{\rm N}$ in Table \ref{linefluxtable}) to an integrated stellar velocity dispersion ($\sigma_*$), we adopt the empirical relations for $z\sim1$ galaxies of \citet{beza18} (as done by other works, e.g. \citealt{uble23}). This results in $\sigma_*\simeq 1.26\times FWHM_{\rm N}/2.355$, or $\sigma_*=144.3\pm2.7$\,km\,s$^{-1}$.

This mass is dependent on three morphological properties ($n$, $q$, $r_{\rm e}$), which we determine from the JWST/NIRCam images using \textlcsc{pysersic} \citep{pash23} and the empirical PSFs. 
Of the 20 available NIRCam images, we exclude those with a weak detection of \UC, as quantified by a best-fit flux that is less than $3\times$ the associated uncertainty (F182M, F090W, F140M, F150W2-F162M, F210M, F250M, F070W). Using the results of the remaining images, we measure the inverse variance-weighted mean aspect ratio ($q=0.32\pm0.13$), S\'ersic index ($n_{\rm s}=4.9 +/- 0.9$), and magnification-corrected effective radius ($r_{\rm e}=0.18\pm0.06$\,kpc). This radius is in agreement with the value derived through a similar analysis by \citet[][$r_{\rm e}=0.165\pm0.020$\,kpc]{koko23} and the mean radius of a sample of more than 200 LRDs by \citet[][$0.21\pm0.03$\,kpc in F444W]{zhan25c}. Adopting our best-fit mean NIRCam morphological parameters, we estimate $\log_{10}(M_{\rm dyn}/M_{\odot})=9.6^{+0.1}_{-0.2}$. 


\subsection{Ly$\alpha$ properties}

As noted in previous works (e.g., \citealt{koko23,fuji24}), \UC is a powerful \lya emitting galaxy (LAE). The presence of another nearby LAE (within a projected distance of $<500$\,kpc and redshift difference $\Delta z\sim0.01$ or $\Delta v\sim300$\,km\,s$^{-1}$) suggests that both LAEs reside in the same ionised bubble \citep{fuji24}. We recover a lower \lya flux ($2.60\pm0.23\times10^{-18}$\,erg\,s$^{-1}$\,cm$^{-2}$) and equivalent width ($91\pm20\angstrom$) than previous results (e.g., \citealt{koko23}), but the resulting emission still meets the most conservative criterion for strong \lya emission used by many studies (i.e, $EW_{\rm 0,Ly\alpha}>75\angstrom$; e.g., \citealt{star10,pent18,jone25a}).

From the R100 spectrum, we are able to estimate the \lya escape fraction. The intrinsic ratio of \lya/\hb$=31.87$ is estimated using our fiducial ISM conditions (see Section \ref{dasec}) and \pyneb. Based on the best-fit fluxes of the narrow \hb and \lya emission (with no dust correction, see Section \ref{dasec}), this results in $f_{\rm esc}^{\rm Ly\alpha}=33\pm7\%$. We note that this commonly applied definition of $f_{\rm esc}^{\rm Ly\alpha}$ (e.g., \citealt{jone25a,shim25}) includes the effect of \lya scattering at multiple scales (e.g, ISM, IGM). If we instead assume a high density of $\log_{10}(n_{\rm e}/{\rm cm}^{-3})=7$, then \pyneb outputs a higher intrinsic ratio of \lya/\hb$=35.49$, resulting in a $<1\sigma$ lower value of $f_{\rm esc}^{\rm Ly\alpha}=29\pm6\%$. We find that \UC falls on the strong correlations between $EW_{\rm 0,Ly\alpha}$-$f_{\rm esc}^{\rm Ly\alpha}$ (e.g., \citealt{begl24,goov24}) and $EW_{\rm 0,Ly\alpha}$-$M_{\rm UV}$ measured by other works (e.g., \citealt{jone25a}), as shown in Appendix \ref{Lya_app}. 

We also note that previous works found a link between O32 (tracing ionisation parameter) and $f_{\rm esc}^{\rm Ly\alpha}$ (e.g., \citealt{naka14,yang17,roy23}, but see also e.g., \citealt{izot20,saxe24}). Our high values (see Section \ref{ism_sec} for O32 details) would appear to support this correlation, lying just beyond the scatter of values from other high-redshift works (e.g., \citealt{roy23}).

Some of the properties of \UC (e.g., compact morphology, high \lya escape fraction, nearby LAEs, high ionization parameter traced by O32) are also seen in the $z\sim8.3$ metal-poor LAE CANUCS-A370-z8-LAE \citep{will25}. But this other source is not an LRD, as it lacks broad Balmer emission or a v-shaped continuum. In addition, it is more metal poor. This highlights the diversity of strong LAEs in the early Universe.

\subsection{Continuum properties}

Based on JWST/NIRCam photometry, this source was classified as an LRD (\citealt{labb25}). One of the defining characteristics of these objects is a v-shaped continuum, with a negative (positive) $F_{\lambda}$ slope in the rest-UV (rest-optical). This was confirmed by the JWST/NIRSpec observations of \citet{gree24}, who found $\alpha_{\rm opt}=0.5\pm0.3$ and $\beta_{\rm UV}=-0.7\pm0.2$ (i.e., the `red' nature is not due solely to emission lines). Our fits of the R100 spectrum return a redder optical slope ($\alpha_{\rm opt}=1.8\pm0.2$) and a similar UV slope ($\beta_{\rm UV}=-0.6\pm0.1$). We note that these slopes meet the LRD criteria of $\alpha_{\rm opt}>0$ and $-2.8<\beta_{\rm UV}<-0.37$ used by classification studies (e.g., \citealt{hain25,koce25}), confirming the LRD nature of \UC. 

Because we fit the R100 spectrum in two parts (i.e., before and after the Balmer limit), there is a discontinuity at $\lambda_{\rm rest}=0.3645\,\mu$m with a red/blue flux ratio of $1.3\pm0.2$ (using the best-fit values and uncertainties of each continuum model). Balmer breaks are commonly observed in LRDs (e.g., \citealt{degr25,furt25,ji25,liu25,wang25}), and this ratio falls within the lower end of the scatter for observed Balmer breaks at high redshift (e.g., \citealt{kuru24,vika24, wang24,tang25}). However, the current data are not able to confirm a Balmer break in \UC.

\section{Discussion}\label{disc_sec}

\subsection{Black hole - host galaxy scaling relations}\label{scalsec}

\begin{figure*}
    \centering
    \includegraphics[width=\textwidth]{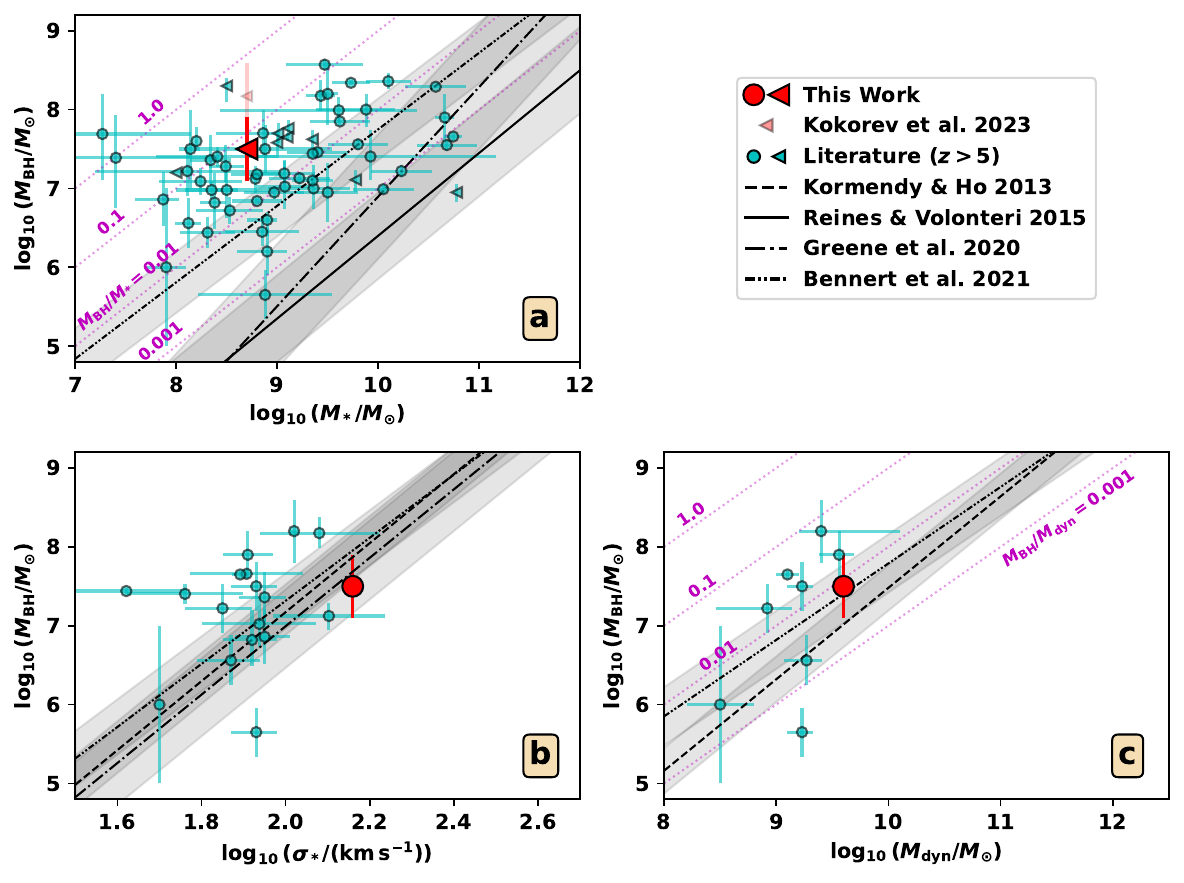}
    \caption{Placement of \UC on planes of $M_{\BH}$-$M_*$ (panel a), $M_{\BH}$-$\sigma_{\rm *}$ (panel b), and $M_{\BH}$-$M_{\rm dyn}$ (panel c, red point). For comparison, we include a sample of $z>5$ AGN studied with JWST (\citealt{goul23,hari23,koce23,koce25,lars23,uble23,chis24,furt24,maio24,maio25_QSO1,chis24,trip24RED,akin25,deug25,juod25,kiyo25,naid25,napo25,rina25,tayl25,zhanJ25}). These values are compared to best-fit relations from local galaxies (\citealt{korm13,rein15,gree20}). The previous estimate of $M_{\BH}$ from \citet{koko23} is included in the $M_{\BH}$-$M_*$ plot.}
    \label{threeplots}
\end{figure*}

Our derived properties allow us to consider the placement of \UC on planes of $M_{\BH}$ as a function of $M_{\rm *}$, $\sigma_*$, and $M_{\rm dyn}$ (Figure \ref{threeplots}), which will inform us on the evolutionary state of the object. In each panel, we compare the placement of \UC relative to other JWST-AGN at $z>5$ (\citealt{goul23,hari23,koce23,lars23,uble23,chis24,furt24,maio24,maio25_QSO1,trip24RED,akin25,deug25,juod25,kiyo25,naid25,napo25,rina25,tayl25,zhanJ25}), as well as correlations found for low-redshift galaxies (\citealt{korm13,rein15,gree20,benn21}).

First, we consider the distribution of black hole mass as a function of stellar mass (Figure \ref{threeplots}a). Most JWST-detected black holes lie above the local relation, reflecting overmassive black holes. Our new $M_{\BH}$ value is lower than that of \citet{koko23}, but it is clear that the black hole of \UC is overmassive, similar to many JWST-detected AGN ($M_{\BH}\gtrsim0.1\times M_{*}$). This is especially evident when comparing to the \citet{rein15} local relation, which is calibrated primarily on AGNs.
Even if our single-epoch black hole mass would be overestimated by $\sim0.7$\,dex, as possibly indicated by the recent results for a highly super-Eddington black hole at $z\sim2.3$ \citep{gravity24}, UNCOVER\_20466 would remain overmassive in the black hole mass -- stellar mass plane. Furthermore, the low Eddington ratio we find in this work suggests that a possible correction towards lower black hole masses may not be as large \citep[see discussion by][]{gravity24}. Additionally, the finding by \citet{juod25_LRD} that the direct BH mass measurement in a LRD at $z=7$ being consistent with local virial relations, supports the idea that the BH mass inferred by us in UNCOVER\_20466 is reasonably accurate.

The relation between black hole mass and galaxy velocity dispersion (also known as the `M-$\sigma$' relation, Figure \ref{threeplots}b) has been found to be constant over a range of redshifts ($z<1$; e.g., \citealt{shen15}). This connection represents a coeval evolution of the black hole and the material surrounding it. As already pointed out by \citet{maio24} and \citet{juod25}, most of the JWST-detected AGN fall onto the local correlation, including \UC. One minor note is that the $\sigma_{\rm *}$ measured from JWST data represents the narrow linewidth from the full source (corrected for instrumental broadening and converted to stellar dispersion), while the same value from local galaxies is extracted only from the bulge. However, most galaxies entering the local relation are ellipticals or bulge-dominated galaxies, i.e. where the bulge contains most of the stellar mass.

The last correlation is between black hole mass and dynamical mass (Figure \ref{threeplots}c). The JWST-detected AGN are scattered on both sides of the local relation, and \UC agrees with the \citet{benn21} relation. 

We note that each of the low-redshift correlations originates from different samples. The $M_{\BH}$-$\sigma_{\rm *}$ and $M_{\BH}$-$M_{\rm dyn}$ correlations of \citet{korm13} are found using a sample of local elliptical and bulge galaxies, where $M_{\rm dyn}=M_{\rm bulge}$. Similarly, \citet{gree20} fit $M_{\BH}$-$\sigma_{\rm *}$ and $M_{\BH}$-$M_{\rm *}$ using a sample of local spiral and elliptical galaxies. The sample used to derive the $M_{\BH}$-$M_{\rm *}$ relation of \citet{rein15} consisted of local broad line AGN. \citet{benn21} combined local broad line AGN and quiescent galaxies to examine all three correlations. Since their data allow for the spatial decomposition of the bulge and disk of each source, $M_{\rm dyn}$ is derived using the radius and velocity dispersion of the central bulge. On the other hand, the observed values for high-redshift galaxies are derived in similar ways.

Figure \ref{threeplots}a shows that \UC has an overmassive black hole (with respect to the stellar mass), especially compared to AGN in the local Universe. Despite this, it already lies on correlations between $M_{\rm BH}$ and bulge properties (velocity dispersion and dynamical mass, Figures \ref{threeplots}b,c). The fact that \UC falls on the same correlations suggests that it may evolve along those relations into a present-day massive early-type galaxy.

\subsection{Black hole growth through cosmic time}\label{cosmictime}

\begin{figure*}
\centering
\includegraphics[width=\textwidth]{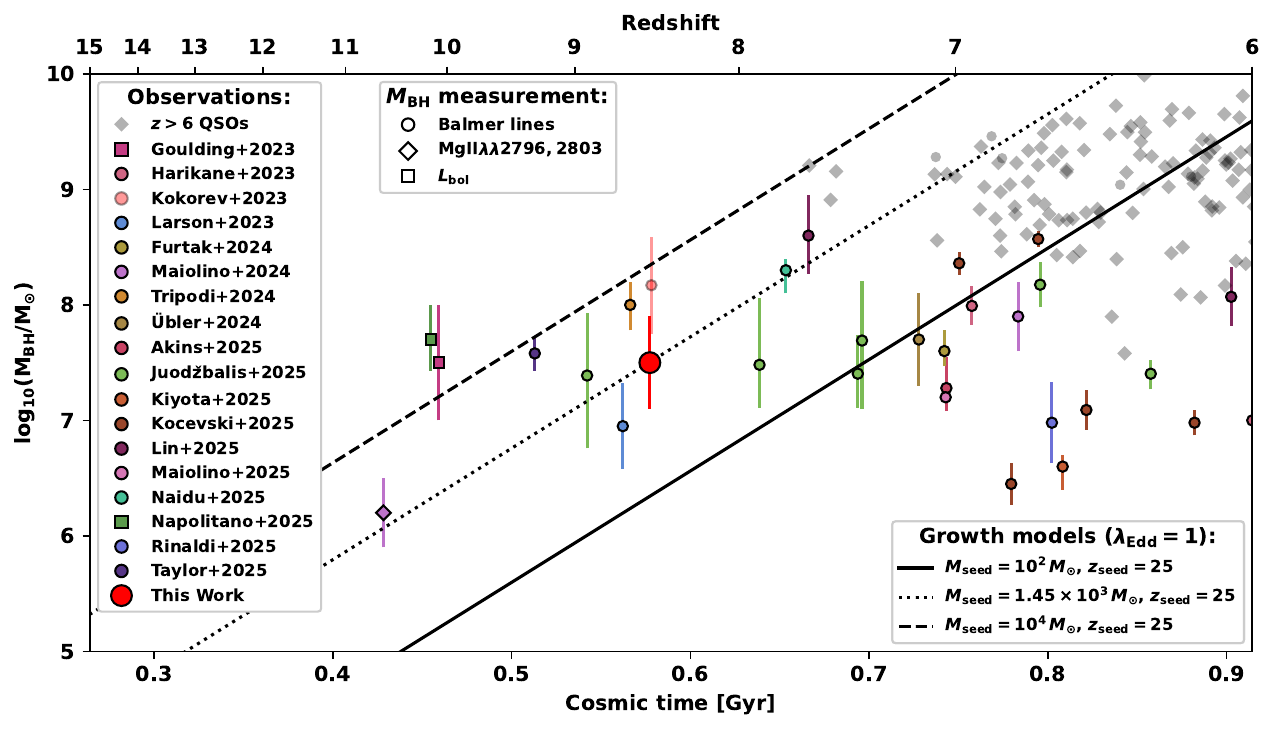}
\caption{Distribution of black hole masses as a function of cosmic time. Our derived mass for \UC is shown by a large red circle, while a faint circle shows the previously determined value of \citet{koko23}. We include a collection of $z>6$ QSOs (\citealt{fan23,mars23,mars24,onou25}) and AGN observed by JWST (\citealt{goul23,hari23,lars23,furt24,maio24,maio25_QSO1,trip24RED,uble24,akin25,juod25,kiyo25,koce25,naid25,napo25,rina25,tayl25}) for comparison. Galaxies where $M_{\rm BH}$ was found through Balmer line (i.e., \ha or \hb) scaling relations are shown by circles, while diamonds show where MgII$\lambda\lambda2796,2803$ scaling relations were used, and studies that used bolometric luminosity scaling relations are shown as squares. Three simple growth models with different seed masses assuming a seeding redshift of 25 and constant Eddington-limited accretion are plotted as black lines.}
\label{bhev}
\end{figure*}

It is clear that \UC features a significant black hole mass at an early cosmic epoch ($t_{\rm H}\sim580$\,Myr). To explore this further, we compare the black hole mass to those of other $z>6$ AGN (Figure \ref{bhev}). 

First, we consider a simple model for how a black hole will grow through cosmic time. Their growth rate may be derived using their Eddington luminosity and Eddington ratio (e.g., \citealt{li12}):
\begin{equation}
\dot{M_{\rm BH}} = \frac{\lambda_{\rm Edd} L_{\rm Edd}}{\epsilon_oc^2}
\end{equation}
where $\epsilon_o$ is the efficiency at which mass is converted to luminosity (here assumed to be 0.1; e.g., \citealt{bosm25}). If each black hole has mass $M_{\rm seed}$ at cosmic time $t_{\rm seed}$ and we assume Eddington-limited accretion ($\lambda_{\rm Edd}=1$), then we may solve the above equation to determine the mass evolution:
\begin{equation}
M_{\rm BH}(t)=M_{\rm seed}e^{(t-t_{\rm seed})/{0.045\,\rm Gyr}}
\end{equation}
We use this equation to calculate three growth curves for seeds at $z=25$: a direct collapse black hole with $M_{\rm seed}=10^4\,M_{\odot}$, a supermassive star with $M_{\rm seed}=10^2\,M_{\odot}$, and a seed tuned to match the observed properties of \UC ($M_{\rm seed}=1.45\times10^3\,M_{\odot}$).

When compared to the literature, \UC is one of the highest-redshift massive black holes, lying close to other sources at $t_{\rm H}\sim0.5-0.6$\,Gyr (i.e., \citealt{lars23,trip24,juod25,tayl25}). The black hole masses of these sources (which were also derived using Balmer line scaling relations) lie between the growth curves for the $10^2\,M_{\odot}$ and $10^4\,M_{\odot}$ seeds. 

The application of $M_{\rm BH}$ scaling relations based on broad Balmer emission is limited by the maximum $\lambda_{\rm obs}$ of the data. Using the standard R100 wavelength limit of $\lambda_{\rm obs}=5.30\,\mu$m, \ha may only be observed up to $z<7.07$, while \hb may be observed up to $z<9.90$. Recent works have successfully extended the wavelength range of JWST/NIRSpec data to $\lambda_{\rm obs}\lesssim5.5\,\mu$m (e.g., \citealt{deug25,donn25}; see also Dawn JWST Archive\footnote{\url{https://dawn-cph.github.io/dja/}} v4 NIRSpec data release), which allows for \ha and \hb detection to higher redshifts. Balmer lines at even higher redshifts can be explored using the broad wavelength coverage of JWST/MIRI (e.g., \citealt{prie25}). While lower-order Balmer lines (e.g., \hg, \hd) are theoretically observable at higher redshift, their intrinsic faintness relative to \ha and \hb make their detection difficult.

Instead, the black holes of higher-redshift galaxies may be measured in other ways. Three $z>10$ AGN have been characterised, where two of them feature similar black hole masses to \UC (\citealt{goul23,napo25}). Due to the lack of observed Balmer lines, these masses were estimated using $L_{\rm bol}$ relations. In addition, both are $X$-ray luminous, suggesting a different nature than LRDs. Finally, the currently highest-redshift black hole (GN-z11; \citealt{maio24}) was estimated using a MgII scaling relation, but falls on the same growth curve as \UC. 

\subsection{Evolution of \UC}

Here, we synthesise the previous subsections to consider the past and future evolution of \UC. The correlation plots of Section \ref{scalsec} show that \UC features a black hole that is overmassive compared to its stellar mass, while it lies on the $M_{\BH}-\sigma_*$ and $M_{\BH}-M_{\rm dyn}$ relations. Combined with the significant broad component of \hb ($FWHM_{\rm B, Ba}\sim2500$\,km\,s$^{-1}$) and low Eddington ratio ($\sim10\%$), this can be interpreted as evidence for rapid black hole accretion in the past (accompanied by low SFR, or small change in M$_*$) which has declined. 

Through a comparison of the \UC black hole mass to model and literature values, we find that while our $M_{\rm BH}$ for \UC is lower than previously found \citep{koko23}, it still requires a heavy seed (i.e., $M_{\rm seed}>10^3\,M_{\odot}$) at high redshift, assuming constant Eddington-limited accretion. Of course, a constant Eddington-limited accretion over nearly 600\,Myr is not physical. The actual mass accretion rate will likely vary over time, with both sub- and super-Eddington periods (e.g., \citealt{li12,hu25,Schneider2023,Trinca2024}). Simulations present a range of such accretion rates, with some showing that super-Eddington accretion could be sustained for up to tens of Myr (e.g., \citealt{lupi24,prol25}). But our constraint on the current Eddington ratio of \UC does not allow us to comment on the past accretion.

Regardless, the high black hole mass of \UC and similar high-redshift AGN are more easily explained if they evolved from a heavy seed. The high mass could also be explained by a primordial black holes (PBHs), which models predict to feature high masses ($\gtrsim10^3\,M_{\odot}$ at $z\sim25$; e.g., \citealt{daya24,zhan25,zhan25b,zipa25}).

The future evolutionary track of \UC is non-trivial to determine, as we only have a constraint on the mass of the black hole and limits on the stellar and warm dust components. If there is a reservoir of molecular gas, then one possibility is that the black hole accretion remains low (as evidenced by the low $\lambda_{\rm Edd}$), while the gas is turned into stars. This would allow \UC to stay on the $M_{\rm BH}-M_{\rm dyn}$ and $M_{\rm BH}-\sigma_{\rm *}$ relations of Figure \ref{threeplots}c and \ref{threeplots}b, while moving closer to the $M_{\rm BH}-M_{\rm *}$ relation of Figure \ref{threeplots}a. However, the gas reservoir of this galaxy is not yet explored, and future ALMA observations are required. We note that the non-detection of \cii or FIR emission in some LRDs \citep{xiao25} suggests a small molecular gas reservoir. Alternatively, this system may have a substantial dark matter contribution, and later baryonic accretion may bring it closer to the local $M_{\rm BH}-M_{\rm *}$ relation \citep{McClymont2026}.

However, the current black hole accretion rate is non-zero (i.e., $\lambda_{\rm Edd}\sim 0.1$), so the black hole mass will increase with cosmic time. One extreme possibility is shown by the dotted line of Figure \ref{bhev}, which shows how $M_{\rm BH}$ would change with Eddington-limited accretion. This curve extends into the high-mass region of $z>7$ QSOs \citep{fan23}, but a more physical sub-Eddington (or variable) growth would result in a lower mass, more similar to other observed LRDs at $z\sim6-8$. 


Together, we find that \UC likely evolved from a heavy seed at high redshift, featured strong AGN activity in the past, and at will either evolve by $z\sim6$ into a source similar to LRDs (assuming low or negligible black hole growth) or QSOs (assuming a high growth rate). Recent studies have found a sample of $z<4$ galaxies with red inner regions and blue star-forming outskirts, either representing descendants of LRDs that have gained a blue envelope over time \citep{bill25} or showing that LRDs may already have faint star-forming regions that are simply not detectable at high redshift \citep{rina25_sag}. 
This represents a possible picture of \UC, but there are several questions that are not yet answered. First, we cannot yet constrain the formation mechanism for the ionised bubble (see discussion of \citealt{fuji24}) that results in a high \lya escape fraction and equivalent width. This could be caused by significant UV emission from young stars in the past, or by AGN activity (e.g., \citealt{wits24,jian25}). In addition, the dust geometry in the host galaxy and around the central black hole is not constrained. We also note the extremely large \oiiic/\oiiib ratio suggests either abnormally high density or temperature. \UC represents one of the most extreme sources at high redshift, and requires additional observations to further constrain its properties (e.g., with ALMA to constrain gas properties or JWST/NIRSpec G140M/F070LP to spectrally resolve the \lya emission).
 
\section{Conclusions}\label{conc_sec}

In this work, we presented new JWST/NIRSpec IFU data of the $z=8.50$ source \UC, taken as part of the BlackTHUNDER survey. While this source is spatially compact ($R_{\rm e}\lesssim200$\,pc), our data enable accurate spectral extraction using a PSF-defined aperture loss correction verified by comparison with JWST/NIRCam photometry. By extracting integrated spectra of both the PRISM/CLEAR and G395H/F290LP data cubes and fitting each, we are able to measure continuum and line properties, yielding constraints on the properties of this source.

\begin{itemize}

\item We confirm the LRD status of \UC, with a v-shaped continuum ($\beta_{\rm UV}=-0.7\pm0.1$, $\alpha_{\rm opt}=1.8\pm0.2$) and compact morphology ($r_{\rm e}=0.18\pm0.06$\,kpc, based on fits to NIRCam images). Similarly to many other LRDs, this source features broad Balmer emission ($FWHM=2503\pm176$\,km\,s$^{-1}$) and is FIR-faint (e.g., \citealt{case25,fuji23}).

\item The best-fit Balmer decrements are in agreement with the intrinsic values, which suggests a lack significant dust attenuation. The previous detection of higher attenuation ($A_{\rm V}\sim2$; \citealt{koko23}) is discussed, including possible reasons for the disagreement with our fiducial values of $A_{\rm V}\sim0$.

\item The broad \hb emission and underlying continuum was used to constrain the black hole properties ($M_{\BH}\sim10^{7.5-7.6}\,M_{\odot}$, $\lambda_{\rm Edd}\sim10\%$), which are slightly lower than previous estimates ($<2\sigma$ discrepant). Using an upper limit on $M_{*}$ from literature, we confirm that \UC features an overmassive black hole.

\item Our observed line ratios are combined with empirical relations and \textlcsc{pyneb} to explore the properties of the host galaxy or gaseous envelope. We find a low \oiia/\oiib ratio and a high \oiiic/\oiiib ratio, which could be explained by a multi-phase ISM (i.e., diffuse gas and dense gas) or an exceptionally high temperature ($T_{\rm e}\gtrsim 10^5$\,K). Other line ratios reveal a low gas-phase metallicity of $0.12\pm0.02$ solar (from high-redshift line ratio calibrations of SFGs, with a higher value from AGN diagnostics) and a high ionisation parameter of $\log_{10}(U)=-1.3\pm0.3$. While the applicability of these relations to high-redshift LRDs is not proven, this high temperature, high-ionisation, extreme environment aligns with the compact AGN nature of this source.

\item By leveraging an extensive grid of photoionization models, we have found that the extremely high \oiiic/\hg ratio indicates that the gas in the host galaxy is not only ionized and over-heated by the AGN, but it also requires extremely high densities, of about $10^7$\,cm$^{-3}$. Together with the very high redshift at which this object is found (being one of the most distant LRDs), this finding is suggestive of a black hole that is forming in an ultra-dense protogalaxy in the early Universe.

\item \lya emission is significantly detected, and is used to estimate an escape fraction of $f_{\rm esc}^{\rm Ly\alpha}\sim30\%$. This places \UC on existing correlations of $EW_{\rm 0,Ly\alpha}$-$f_{\rm esc}^{\rm Ly\alpha}$ and $EW_{\rm 0,Ly\alpha}$-$M_{\rm UV}$.

\item While \UC features an overmassive black hole, we find that it lies on local correlations of $M_{\rm BH}-\sigma_{*}$ and $M_{\rm BH}-M_{\rm dyn}$. Since these correlations were primarily determined using the properties of local bulges, the observable emission of \UC may represent the progenitor of the central bulge of a lower-redshift galaxy.

\item We compare the black hole mass of \UC to other $z>6$ AGN. While our new $M_{\rm BH}$ is lower than previous estimates, we find that it is still one of the highest-redshift and most massive black holes. Using  a set of growth tracks assuming Eddington-limited accretion, we show that this source would likely require a heavy seed ($M_{\rm seed}\gtrsim10^3$ assuming $z_{\rm seed}\sim25$). 

\item The combination of properties is synthesised to predict the history and future evolution of this source. One possibility is that \UC underwent a period of rapid black hole accretion in the recent past (which increased $M_{\rm BH}$ and $\sigma_{\rm *}$ of the host galaxy) which has now ended (as evidenced by the low $\lambda_{\rm Edd}$). Depending on future accretion, it may evolve into a black hole similar to those of QSOs (high accretion) or LRDs (negligible accretion) at $z\sim6$.

\end{itemize}

Our data show that \UC is an exceptional source in the early Universe, with typical LRD properties, an overmassive black hole, evidence for a large ionised bubble, and extreme ISM conditions. Additional observations at different wavelengths (e.g., JWST/MIRI, ALMA [CII]$158\mu$m) will help to reveal the complex nature of this source. 

\section*{Acknowledgements}

We thank the anonymous referee for constructive feedback that has enhanced this work.
GCJ, RM, XJ, FDE, IJ, and JS acknowledge support by the Science and Technology Facilities Council (STFC), by the ERC through Advanced Grant 695671 ``QUENCH'', and by the UKRI Frontier Research grant RISEandFALL.
H\"U and GM acknowledge funding by the European Union (ERC APEX, 101164796). Views and opinions expressed are however those of the authors only and do not necessarily reflect those of the European Union or the European Research Council Executive Agency. Neither the European Union nor the granting authority can be held responsible for them.
RM also acknowledges funding from a research professorship from the Royal Society.
AJB acknowledges funding from the ``FirstGalaxies" Advanced Grant from the European Research Council (ERC) under the European Union’s Horizon 2020 research and innovation programme (Grant agreement No. 789056).
GC acknowledges support from the INAF GO grant 2024 “A JWST/MIRI MIRACLE: MidIR Activity of Circumnuclear Line Emission”.
KI acknowledges support from the National Natural Science Foundation of China (12233001), the National Key R\&D Program of China (2022YFF0503401), and the China Manned Space Program (CMS-CSST- 2025-A09).
P.G.P.-G and MP acknowledge support from grants PID2022-139567NB-I00, PID2024-159902NA-I00, and RYC2023-044853-I funded by Spanish Ministerio de Ciencia e Innovaci\'on MCIN/AEI/10.13039/501100011033, FEDER {\it Una manera de hacer Europa}, and El Fondo Social Europeo Plus FSE+.
RS acknowledges support from the PRIN 2022 MUR project 2022CB3PJ3—First Light And Galaxy aSsembly (FLAGS) funded by the European Union—Next Generation EU, and from EU-Recovery Fund PNRR - National Centre for HPC, Big Data and Quantum Computing.
ST acknowledges support by the Royal Society Research Grant G125142.
This work is based in part on observations made with the NASA/ESA/CSA James Webb Space Telescope. The data were obtained from the Mikulski Archive for Space Telescopes at the Space Telescope Science Institute, which is operated by the Association of Universities for Research in Astronomy, Inc., under NASA contract NAS 5-03127 for JWST.

\section*{Data Availability}
 
The NIRSpec data used in this research was obtained within the NIRSpec-IFU GO programme `Unveiling the nature and impact of the first population of black holes: an extensive NIRSpec-IFU survey in the first billion years' (PID 5015; \url{https://doi.org/10.17909/m9xt-6908}) and is publicly available. Data presented in this work will be shared upon reasonable request to the corresponding author.



\bibliographystyle{mnras}
\bibliography{example} 





\appendix

\section{IFU-MSA spectral comparison}\label{ifumsa}
As part of our analysis of the JWST/NIRSpec IFU observations of \UC, we compare the best-fit properties to the JWST/NIRSpec MSA observations of this source as presented by \citet{koko23}. While many properties are the same (e.g., redshift, presence of broad \hb and an overmassive black hole), some line fluxes are significantly different (up to a factor of $\sim2$; see Table \ref{linefluxtable}). Here, we investigate the origin of this discrepancy.

\begin{figure*}
    \centering
    \includegraphics[width=\textwidth]{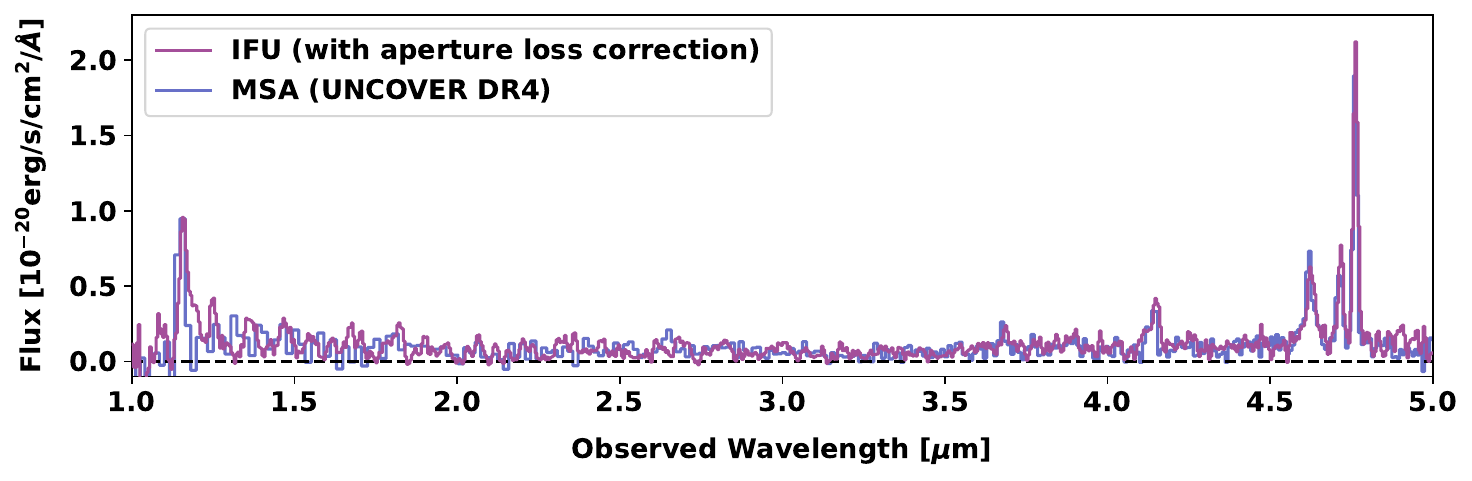}
    \caption{\UC spectra from this work (pink) compared to NIRSpec MSA observations \citep[][blue]{pric25}. All data have been corrected for gravitational magnification and slit or aperture loss.}
    \label{ifumsafig}
\end{figure*}

First, we consider differences in the studied spectra, through extraction method or aperture/slit loss correction. The calibrated UNCOVER MSA spectrum (\citealt{pric25}; available from the survey website\footnote{\url{https://jwst-uncover.github.io/DR4.html}}) is compared to our extracted IFU spectrum (with aperture loss correction) in Figure \ref{ifumsafig}. While the two spectra feature small differences, there are no large-scale or wavelength-dependent differences between them. 

Instead, we consider our treatment of the NIRSpec LSF. While we use the fiducial LSF directly, \citet{koko23} assume LSF/LSF$_{\rm fiducial}=0.59$ \citep{degr24}. If we instead adopt  this narrower LSF, then we find slightly smaller intrinsic linewidths and larger fluxes for some lines (but still not in agreement with the values of \citealt{koko23}). If we allow the deviation from the LSF to vary, then we find a best-fit value of LSF/LSF$_{\rm fiducial}=0.89$ and similar line fluxes as the fiducial case. Therefore, we do not consider the LSF treatment to be responsible for the discrepancy with previous results, and adopt the fiducial value in our analysis.

As a final test, we consider the possibility that our use of a Gaussian profile is to blame for the discrepancy in fluxes. We estimate a non-parametric total flux of \oiiib by isolating the spectral range around this line ($4.73<\lambda_{\rm obs}/\mu \rm m<4.78$), subtracting the best-fit continuum flux, and integrating the IFU spectra - resulting in a slightly lower value than the best-fit integrated flux ($324\times 10^{-20}$ erg s$^{-1}$ cm$^{-2}$ \textit{vs.} $\sim360\times 10^{-20}$ erg s$^{-1}$ cm$^{-2}$ in Table \ref{linefluxtable}).

These tests show that the UNCOVER NIRSpec MSA spectrum agrees with the NIRSpec IFU spectrum studied here, and that the higher fluxes in a previous work are not due to the assumed LSF or assumed line shape. Both spectra were corrected by the same gravitational magnification, and comparisons to NIRCam data have confirmed that their slit or aperture loss corrections are valid. While the discrepancy may be caused by a combination of multiple effects, we note that due to the conservative uncertainties presented by \citet{koko23}, the redshift, FWHMs and most line fluxes agree to within $3\sigma$ (e.g., \lya, \oiiab, \neiiia). 

\section{JWST/NIRCam filters}\label{filterapp}
\UC benefits from a wealth of JWST/NIRCam imaging from two programs (see Section \ref{NS_sec}). We use a subset of these images to test the flux calibration of our data (Section \ref{NCNSsec}), and to extract morphological properties of \UC (Section \ref{mdynsec}). The filters of use are listed in Table \ref{NC_table}, along with their program of origin, pivot wavelength\footnote{Pivot wavelength taken from JDOX: \url{https://jwst-docs.stsci.edu/jwst-near-infrared-camera/nircam-instrumentation/nircam-filters}}, and empirical PSF FWHM\footnote{NIRSpec PSFs taken from JDOX: \url{https://jwst-docs.stsci.edu/jwst-near-infrared-camera/nircam-performance/nircam-point-spread-functions}} in Table \ref{NC_table}.


\begin{table}
\centering
\begin{tabular}{c|ccc}
Filter	&	PID	&	$\lambda_{\rm pivot}$ 	&	$FWHM_{\rm PSF,emp}$	\\
	&		&	[$\mu$m]	&	[$''$]	\\ \hline
F335M	&	4111	&	3.362	&	0.111	\\
F356W$\dagger$	&	2561	&	3.565	&	0.116	\\
F360M	&	4111	&	3.623	&	0.120	\\
F410M$\dagger$	&	2561	&	4.083	&	0.137	\\
F430M	&	4111	&	4.281	&	0.144	\\
F444W$\dagger$	&	2561	&	4.402	&	0.145	\\
F460M$\dagger$	&	4111	&	4.630	&	0.157	\\
F480M$\dagger$	&	4111	&	4.817	&	0.164	\\ \hline
\end{tabular}
\caption{Details of JWST/NIRCam filters used in this work. $\dagger$Fit with \textlcsc{pysersic} to find morphological properties (Section \ref{mdynsec}).}
\label{NC_table}
\end{table}

\section{Serendipitous source}\label{sere_sec}
Despite the small field of view of our NIRSpec IFU data ($\sim3.6''\times3.8''$ due to dithering and drizzling), we also find a serendipitous line-emitting source which we name `BlackBolt-1' (inspired by the theme of `BlackTHUNDER'). This source lies on the edge of the field of view, $\sim2.2''$ to the north-west of \UC (i.e., at 00:14:33.614 -30:23:09.4467). This location places it within $0.2''$ of an object in the `Medium Bands, Mega Science' data release\footnote{\url{https://jwst-uncover.github.io/DR3.html}} (ID 21646; \citealt{sues24}) which has $z_{\rm phot}=6.68^{+0.03}_{-0.04}$ based on the JWST/NIRCam photometry. 

Because of its location, we are not currently able to present a well-calibrated spectrum (i.e., some data are excluded due to high noise levels, and the PSF is not well determined at the edge of the field of view). However, we are able to define an aperture by performing a 2-D Gaussian fit to the emission of the brightest spectral channel and taking the resulting best-fit major axis FWHM ($0.2''$). By extracting a spectrum from the R100 cube centred on this location using a circular aperture of radius $0.2''$, we may determine its spectroscopic redshift.

\begin{figure*}
    \centering
    \includegraphics[width=\textwidth]{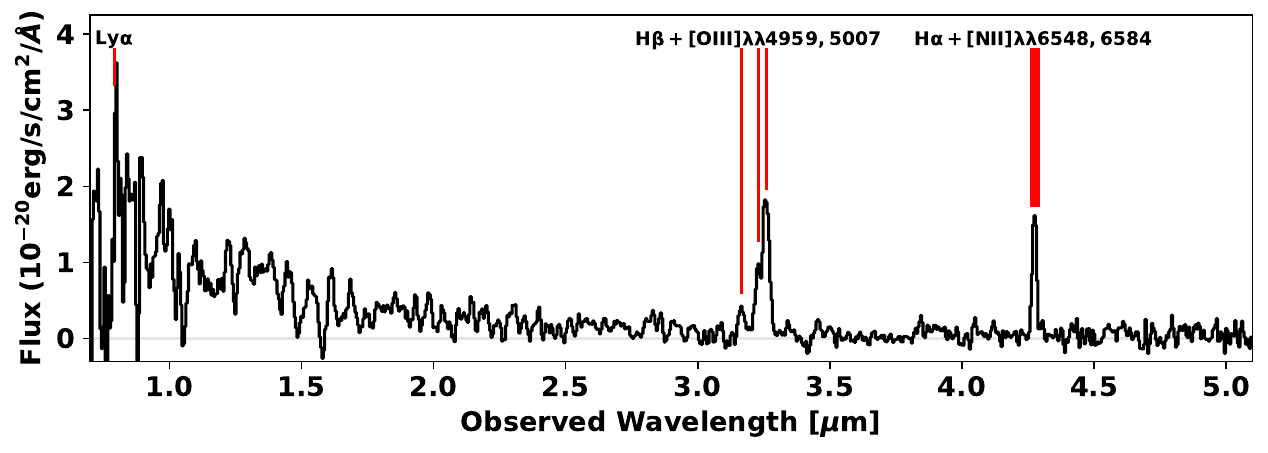}
    \caption{Integrated spectrum for a serendipitous source identified within the field of view (BlackBolt-1), extracted from 00:14:33.614 -30:23:09.4467 using a circular aperture of radius $0.2''$ from the R100 cube. No corrections have been applied, and the line fluxes are affected by noise artifacts due to its location on the edge of the data cube. Despite this, we identify the spectroscopic redshift of $z_{\rm spec}=5.511$ based on several line detections (red lines).}
    \label{bb1}
\end{figure*}

As seen in Figure \ref{bb1}, BlackBolt-1 features at least four strong line detections, all at different observed wavelengths than the lines detected from \UC (i.e., this is not a lensed image). We use these lines to identify the redshift to be $z_{\rm spec}=5.511$, based on the identification of the [OIII]-H$\beta$ complex and \ha, as well as the \lya break.

Since this source is at a much lower redshift than our target, it is observed at a very different epoch (i.e., $\sim1.0$\,Gyr after the Big Bang, rather than $\sim0.58$\,Gyr for \UC). With this in mind, we do not include BlackBolt-1 in our analysis, but note that further analysis may be performed in a future work. 

\section{Details of aperture loss correction}\label{ALC_sec}

When analysing the integrated spectra of \UC, we choose an extraction aperture of radius $0.125''$ in order to maximise the S/N of \oiiib. This aperture captures most of the emission from this compact object, but some emission from the wings of the PSF is missed. This fraction of missed flux is not constant, as the NIRSpec IFU PSF size changes with wavelength. In order to recover the total flux, an aperture loss correction is required (see similar analyses for NIRSpec IFU data in \citealt{arri24,zamo25}). In this Appendix, we detail how we calculate wavelength-dependent aperture loss corrections for our extracted spectra.

\subsection{NIRSpec IFU PSF derivation}\label{psfsec}

We begin by analysing the shape of the NIRSpec IFU PSF. An in-depth characterisation of the NIRSpec PSF was performed by \citet{deug24}. This work found a strong evolution of the PSF FWHM with wavelength, as well as a significant difference in the PSF width across and along the NIRSpec slice directions (up to $\sim40\%$). Although three approaches were utilised, no concrete PSF shape was found. The results of this work were adopted by many other NIRSpec/IFU studies (e.g., \citealt{marc24,vent24}). To verify these results, we analyse archival observations of calibration stars, and compare the results to empirical models.

Observations of stars were included in programs in order to calibrate the PSF of NIRSpec (PSFSTARUHSJ0439 in PID 1222, PI C. J. Willott; PSFSTARULASJ1342 in PID 1219, PI N. Luetzgendorf). Additionally, we consider the calibration data of PID 1537 (PI: K. D. Gordon), who observed the white dwarf G191-B2B with the JWST/NIRSpec IFU with all three R1000 gratings and all three R2700 gratings. For all of these observations, we downloaded the calibrated (stage 3) data cubes from the MAST archive and fit the stellar emission in each spectral channel of each cube with a 2D Gaussian model. The geometric mean of the best-fit semi-major axes were recorded, resulting in curves of PSF FWHM as a function of wavelength.

\begin{figure*}
    \centering
    \includegraphics[width=\textwidth]{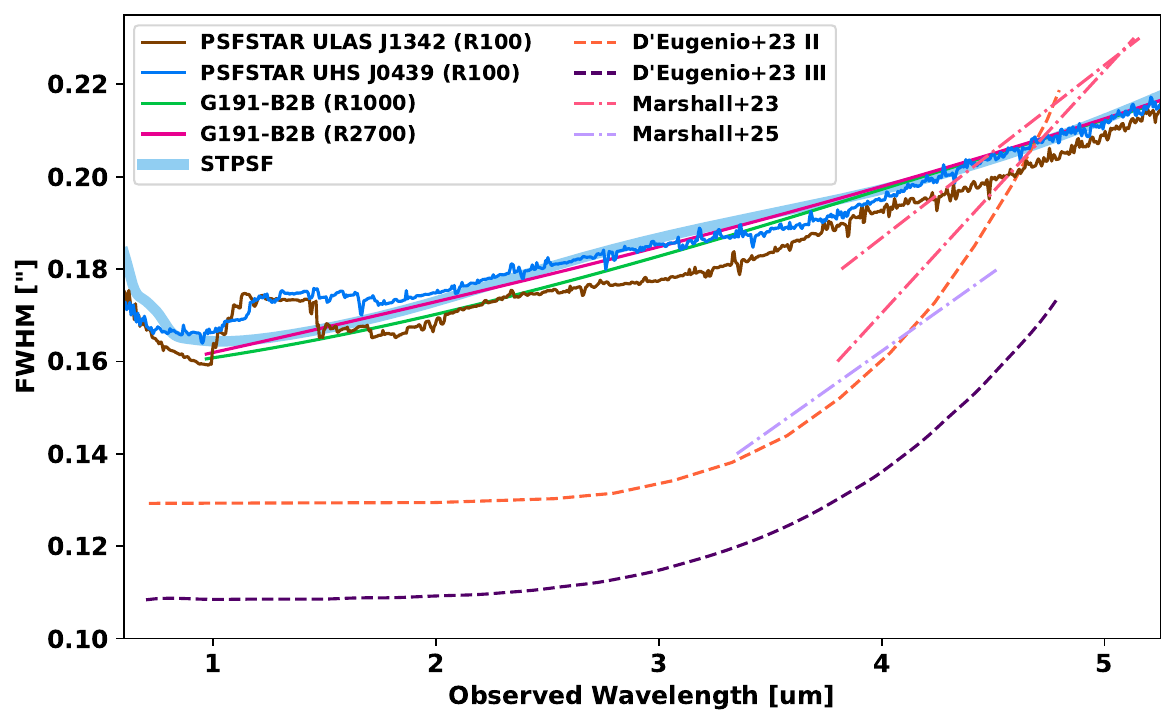}
    \caption{Best-fit FWHMs of the NIRSpec IFU PSF, derived using a variety of methods. Two methods of \citet{deug24} are shown by dashed lines, while the results of \textlcsc{QDeblend3D} fits to QSO observations are shown by dot-dashed lines (\citealt{mars23,mars25}). The FWHM curves derived from observations of three stars are shown by thin lines, while the empirical curve from \textlcsc{stpsf} is depicted with a thick line.}
    \label{PSFFWHMFIG}
\end{figure*}

The resulting curves are shown in Figure \ref{PSFFWHMFIG}. 
Because the data of G191-B2B features significant oscillations in FWHM (`wiggles', see \citealt{pern23,dumo25}), we fit a third-order polynomial to each of the curves for this source. 
PSFSTARULASJ1342 and PSFSTARUHSJ0439 show elevated flux between $\lambda_{\rm obs}\sim1.0-1.5\,\mu$m. This may be due to saturation of the NIRSpec detector, which would result in a non-Gaussian source and thus an inflated FWHM.
For comparison, we also display results from \citet{deug24} (methods II and III), and the FWHM values derived from QSO observations \citep{mars23,mars25} using the code \textlcsc{QDeblend3D} (\citealt{huse13,huse14}).

It is clear that the curves of \citet{deug24} fall beneath those of the calibration star observations. This could be due to the fact that the serendipitous source studied by \citet{deug24} fell on the edge of the field of view rather than the pointing centre. Indeed, their analysis of a separate stellar observation (Method I) resulted in a larger PSF, in better agreement with the stars studied here. 

Finally, we compare all of our approaches to the empirical curve outputted by \textlcsc{stpsf}\footnote{\url{https://stpsf.readthedocs.io/en/latest/}}, the successor to \textlcsc{webbpsf} \citep{perr15}. This code takes many properties of the NIRSpec instrument into account (e.g., geometric distortion, detector charge transfer effects), resulting in physical models of the wavelength dependence of the PSF. The resulting PSF FWHM curve is in agreement with the results of calibration star observations, so we will use this model (hereafter referred to as the fiducial PSF) for our analyses. 

\subsection{Comparison with observed curve of growth}
While this fiducial PSF is in good agreement with the shapes of observed point sources, we note that Appendix \ref{psfsec} and other works (e.g., \citealt{deug24}) used a two-dimensional Gaussian to model the PSF. In reality, the JWST/NIRSpec PSF is more complex, with a bright core and a fainter six-pointed `snowflake' pattern of maxima and minima. While a 2D Gaussian model can describe the dominant central emission, the lower-level emission deviates from this axisymmetric model. We explore this by creating curves of growth (CoGs), which show the amount of flux contained within circular apertures of increasing radius.
The slope of these CoGs are directly related to the density of flux: a bright core will result in a rapid rise, while diffuse emission will create a more shallow increase.
We compare the CoGs for the observed \oiiib emission in the R100 and R2700 cubes, our best-fit 2D Gaussian model to the PSF, and the fiducial PSF. 

\begin{figure}
    \centering
    \includegraphics[width=0.5\textwidth]{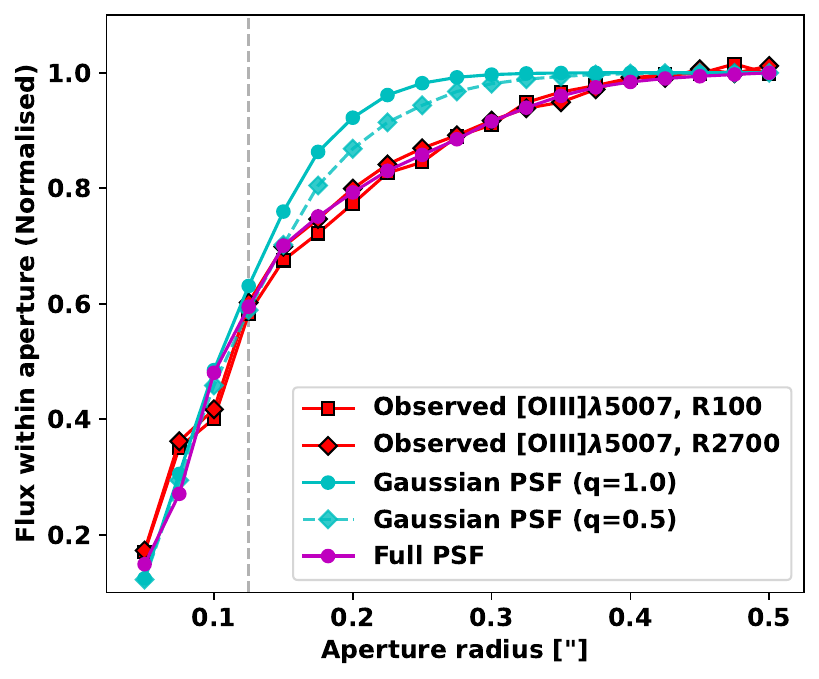}
    \caption{Curves of growth for our observations and models of the fiducial PSF. In red, we show the CoG for \oiiib, as extracted from both observed cubes. The blue lines show the CoG of 2D Gaussian models, assuming different axis ratios ($q$). The purple line shows the CoG of the fiducial model.}
    \label{cog_fig}
\end{figure}

To create the \oiiib CoG, we extracted spectra using circular apertures centred on \UC with a range of radii ($0.025''-0.500''$), isolated the wavelength range around \oiiib, fit the continuum and line emission in the resulting spectrum, and recorded the best-fit line flux for each aperture. The resulting CoGs are shown as red lines in Figure \ref{cog_fig}. In order to create a similar CoG for the fiducial PSF of Section \ref{psfsec}, we integrate the PSF within a range of apertures and normalise each flux to the value in the largest aperture ($r=0.5''$; see magenta line with circular markers in Figure \ref{cog_fig}). Finally, we create CoGs for 2D Gaussian models with the same FWHM as the fiducial model, assuming a circular model ($q=1.0$) and an elongated model ($q=0.5$).

From this analysis, we see that all of the CoGs agree at small radii ($r\lesssim0.125''$) and large radii ($r\gtrsim0.4''$), but the 2D Gaussian model of the PSF deviates at intermediate radii ($0.125''\lesssim r\lesssim 0.400''$). This behaviour indicates diffuse emission in the PSF that is not present in the 2D Gaussian model. Because the full PSF follows the observed CoG more closely, we will use this fiducial PSF rather than a Gaussian approximation.

\subsection{Aperture loss correction}\label{alc_sec}
\begin{figure}
\centering
\includegraphics[width=0.5\textwidth]{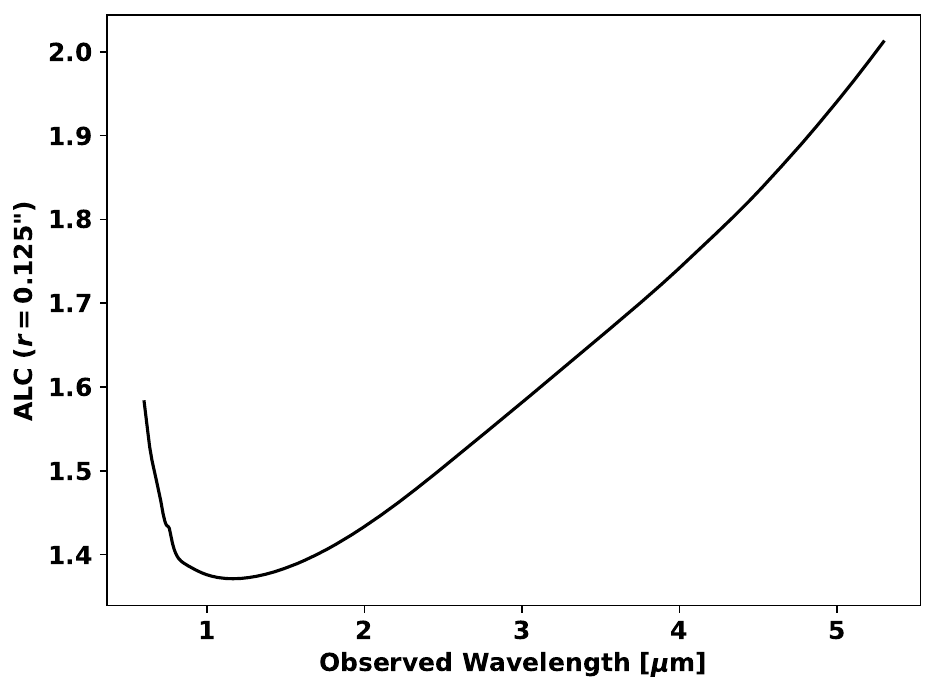}
\caption{Aperture loss correction as function of observed wavelength for a circular aperture with radius $0.125''$. This was derived using the full PSF model from \textlcsc{stpsf}, rather than a Gaussian approximation.}
\label{aclfig}
\end{figure}

We have now characterised the wavelength-dependent FWHM of the NIRSpec IFU PSF, and tested this model on our observations. The remaining step is to derive the aperture loss correction curve for our adopted aperture ($r=0.125''$). This is done by numerically integrating the full corresponding PSF and dividing by the value within the aperture. The resulting correction as a function of wavelength is shown in Figure \ref{aclfig}. 

\subsection{Comparison to NIRCam data}\label{NCNSsec}
While this empirical model is a good match to observations of stars, it is not yet thoroughly tested on observational data of high-redshift galaxies, and thus may not be accurate. To validate this correction, we conduct a comparison of our R100 NIRSpec spectrum and archival JWST/NIRCam photometry of \UC. 

\begin{figure*}
    \centering
    \includegraphics[width=\textwidth]{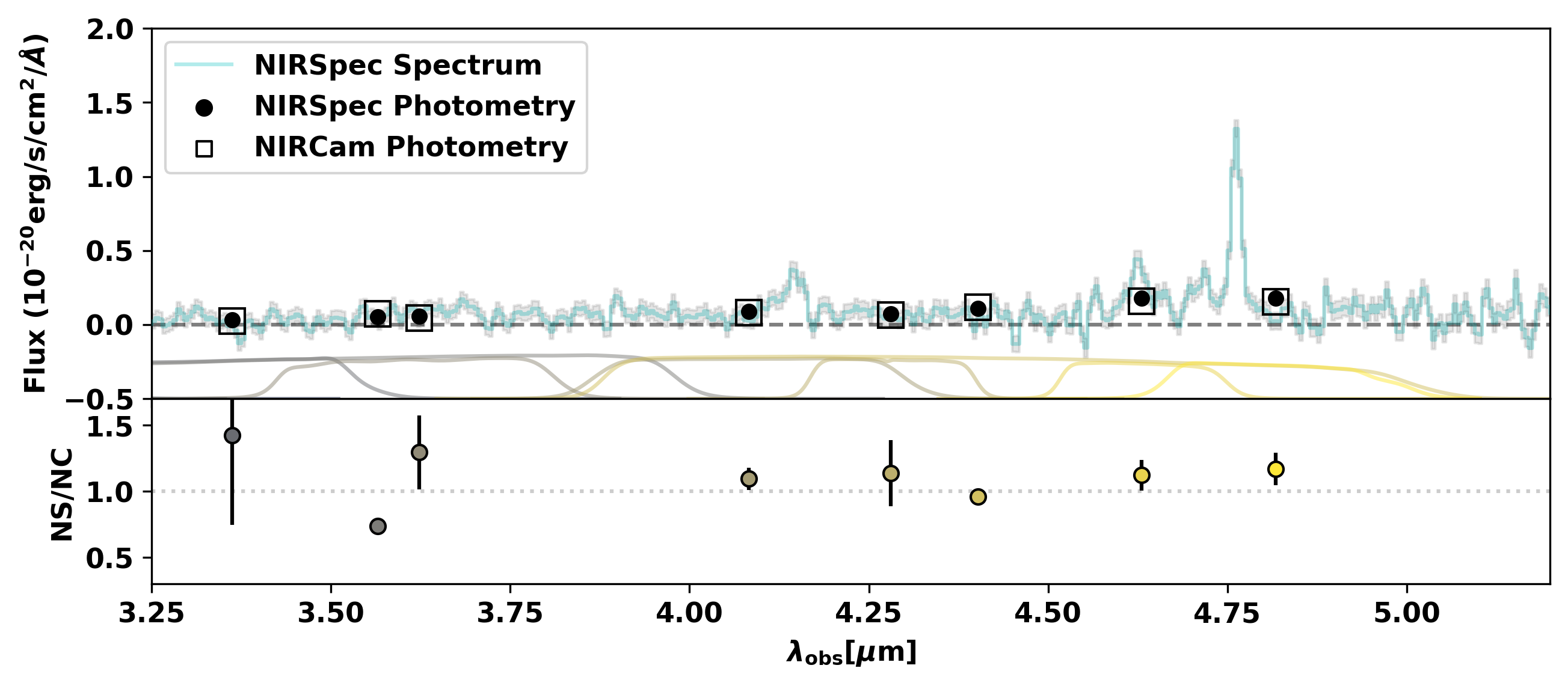}
    \caption{Comparison of the JWST/NIRSpec R100 spectrum and pseudo-photometry with JWST/NIRCam photometry (top panel). The pseudo-photometry was created using the throughput curves for each NIRCam filter, shown in the lower portion of the top panel. This results in ratio of NIRSpec to NIRCam photometric fluxes that are in broad agreement with the expected value from the empirical NIRSpec PSF (lower panel).}
    \label{ncnsfig}
\end{figure*}

First, we extract fluxes from each NIRCam image by fitting each with a 2D circular Gaussian model with a constant offset (to account for background emission), where the FWHM of the model is set to be the fiducial FWHM from JDOX (see Table \ref{NC_table}). In order to compare these photometric points to the NIRSpec data directly, we first convolve the R100 cube with the throughput curve of each NIRCam filter to create a set of pseudo-photometric images, and measure the flux using a circular aperture of $r=0.125''$. We apply the aperture loss correction derived in Appendix \ref{alc_sec} to the NIRSpec fluxes, and compare the resulting values in Figure \ref{ncnsfig}. We limit the comparison to NIRCam filters at $\lambda_{\rm obs}>3\,\mu$m due to elevated noise on the blue side of the NIRSpec data. The resulting fluxes are comparable. 

\subsection{Note on variability}
By comparing JWST/NIRSpec MSA and IFU data taken at different times, \citet{ji25} found that Abell2744-QSO1 featured significant variability. This analysis benefitted from the gravitationally lensed nature of the source: each of the three lensed images represented a different light travel time. Thus, they were able to show that \hb rest-frame equivalent width decreased by $\sim50\%$ over an observer-frame time of $\sim20$\,yr. 

While \UC is not multiply lensed, it is possible that we could use the different data to explore the variability of this source. This is because of the different observation times: the NIRCam data were taken in two epochs (2 November 2022 for PID 2561, 11 November 2023 for PID 4111), while the JWST/NIRSpec MSA data were taken on 31 July 2023 and the JWST/NIRSpec IFU were taken on 5 December 2024. But as demonstrated in Appendix \ref{NCNSsec}, we find good agreement between the NIRCam data taken on each date and the IFU data, and thus no significant evidence for continuum variability on the observer-frame timescale of $\sim2\,$yr. {In addition, the agreement between the MSA and IFU spectra (see Appendix \ref{ifumsa}) suggests no significant change in emission line strength on an observer-frame timescale of $\sim1\,$yr.}

\section{Lyman alpha correlations}\label{Lya_app}

\lya emission in galaxies appears to follow a few well-determined trends, where the \lya equivalent width correlates with the \lya escape fraction (e.g., \citealt{begl24,goov24}), and with $M_{\rm UV}$ (e.g., \citealt{jone24,jone25a}). However, much of this work has been done at lower redshift ($z<8$). While LAEs have been discovered well into the epoch of reionization ($z>10$; e.g., \citealt{wits25}), the high redshift and LRD nature of \UC make it an interesting source to examine in light of these correlations.

\begin{figure*}
    \centering
    \includegraphics[width=\textwidth]{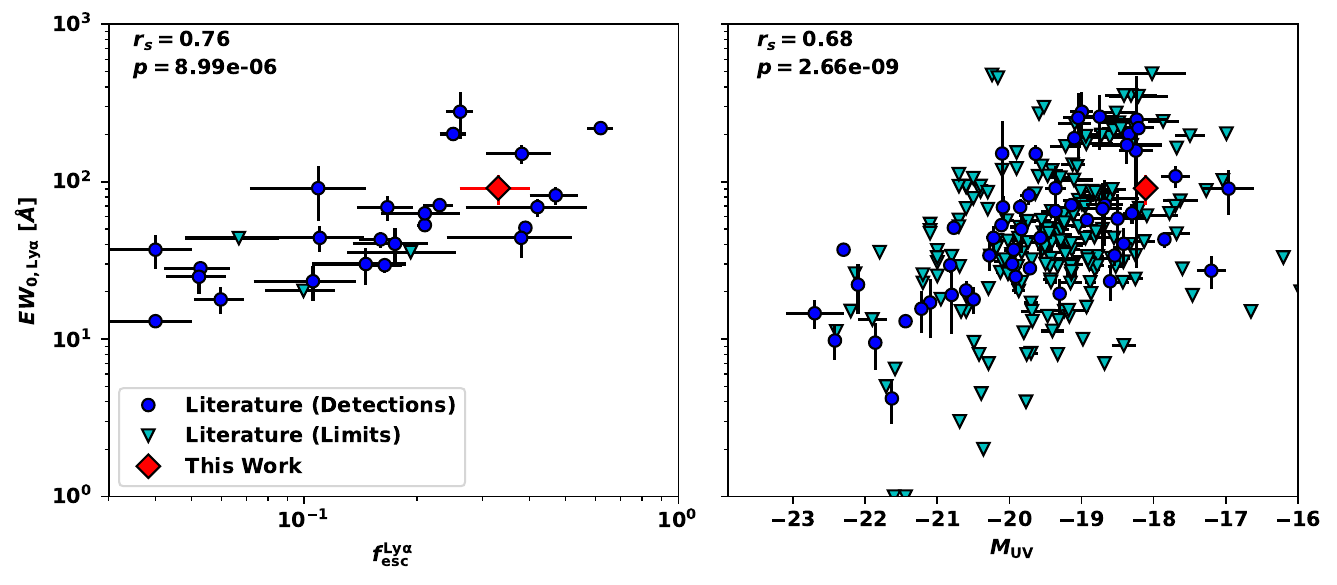}
    \caption{Placement of \UC on distributions of $EW_{\rm 0,Ly\alpha}$ as a function of $f_{\rm esc}^{\rm Ly\alpha}$ (derived with fiducial ISM conditions; left panel) and $M_{\rm UV}$ (right panel), as shown by a red marker. For comparison, we plot the values of $z>7$ galaxies from literature (\citealt{vanz11,ono12,sche12,song16,pent18,hoag19,full20,tilv20,ends22,jung22,tang23,tang24,napo24,jone25a,kage25,will25}. The value of $M_{\rm UV}=-18.11$ for \UC is taken from \citet{fuji24}. For each correlation, we present a Spearman coefficient ($r_s$) and p-value ($p$) for the detections.}
    \label{lyafig}
\end{figure*}

In Figure \ref{lyafig}, we show the placement of \UC on two \lya correlations, and compare it to a set of $z>7$ LAEs from literature. This shows that it falls within the scatter of other similar LAEs, which may be interpreted as evidence as similar avenues of \lya visibility. The Spearman correlation of both distributions are positive with very low p-values, indicating strong positive correlations.

\bsp	
\label{lastpage}
\end{document}